\documentclass[modern]{aastex62}
\usepackage{color}
\usepackage[super]{nth}
\newcommand{\is}{i_{\star}}

\newcommand{\spl}{\delta\nu_{\star}}

\usepackage{soul}
\usepackage{CJK}
\graphicspath{{./}{figures/}}

\received{November 14, 2018}
\revised{February 24, 2019}
\accepted{March 11, 2019}
\shorttitle{Asteroseismic determination of the stellar rotation period}
\shortauthors{Suto, Kamiaka \& Benomar}
\begin{document}
\title{
Asteroseismic determination of the stellar rotation period
of the {\it Kepler} transiting planetary systems
and its implications for the spin-orbit architecture
}
\correspondingauthor{Yasushi Suto}
\email{suto@phys.s.u-tokyo.ac.jp}

\author[0000-0002-4858-7598]{Yasushi Suto}
\affiliation{Department of Physics, The University of Tokyo, Tokyo,
  113-0033, Japan}
\affiliation{Research Center for the Early
  Universe, School of Science, The University of Tokyo, Tokyo
  113-0033, Japan}

\author[0000-0001-6036-3194]{Shoya Kamiaka}
\affiliation{Department of Physics, The University of Tokyo, Tokyo,
  113-0033, Japan}

\author{Othman Benomar} \affiliation{Center for Space Science, NYUAD
  Institute, New York University Abu Dhabi, PO Box 129188, Abu Dhabi,
  UAE}

\begin{abstract}
We measure the rotation periods of 19 stars in the {\it Kepler}
  transiting planetary systems, $P_{\rm rot, astero}$ from
  asteroseismology and $P_{\rm rot, phot}$ from photometric variation
  of their lightcurve. Two stars exhibit two clear peaks in the
  Lomb-Scargle periodogram, neither of which agrees with the seismic
  rotation period. Other four systems do not show any clear peak,
  whose stellar rotation period is impossible to estimate reliably
  from the photometric variation; their stellar equators may be
  significantly inclined with respect to the planetary orbital plane.
  For the remaining 13 systems, $P_{\rm rot, astero}$ and $P_{\rm rot,
    phot}$ agree within 30\%.  Interestingly, three out of the 13
  systems are in the spin-orbit resonant state in which $P_{\rm orb,
    b}/P_{\rm rot, astero} \approx 1$ with $P_{\rm orb, b}$ being the
  orbital period of the inner-most planet of each system. The
  corresponding chance probability is ($0.2$-$4.7$) \% based on the
  photometric rotation period data for 464 {\it Kepler} transiting
  planetary systems. While further analysis of stars with reliable
  rotation periods is required to examine the statistical
  significance, the spin-orbit resonance between the star and planets,
  if confirmed, have important implications for the star-planet tidal
  interaction, in addition to the origin of the spin-orbit
  (mis-)alignment of transiting planetary systems.
\end{abstract}


\vspace*{2cm}

\section{Introduction \label{sec:intro}}

Both diversities and universality exhibited in the observed
architecture of exoplanetary systems should be understood from the
combined outcomes of their initial condition and subsequent
evolution. A puzzling and interesting clue comes from the distribution
of spin-orbit angles of transiting planetary systems.  One may
naturally expect that the spin axis of the host star is well aligned
with the normal vector of the surrounding protoplanetary disk. Since
planets subsequently form within the disk, their orbital axis is
supposed to be parallel to the stellar spin axis, as is exactly the
case for the Solar system.

Measurements of the projected spin-orbit angle, $\lambda$, via the
Rossiter-McLaughlin (RM) effect
\citep{Rossiter1924,McLaughlin1924,Queloz2000,Ohta2005,Winn2005},
however, indicate that 28 out of 124 transiting close-in gas-giant
planets are misaligned in a sense that their $2\sigma$-lower limits of
$\lambda$ exceed $30^\circ$ \citep[see Figure 12 of][]{Kamiaka2019}.
\citet{Albrecht2012} found that those misaligned planets
preferentially orbit around hot central stars with the effective
temperature $T_{\rm eff}>6100$K, and suggested a realignment process
due to the stronger tidal interaction with a thicker convective layer
for cooler stars.

This interpretation implies that the well-aligned initial condition is
significantly broken via the violent dynamical evolution of planets in
late stages, and the inner-most planet becomes realigned toward the
stellar spin axis through the tidal interaction preferentially in cool
host-star systems.  For instance, \citet{Nagasawa2008} showed that
planet-planet scattering and the subsequent Kozai-Lidov effect in
multi-planetary systems significantly modify the orbital inclination
of the inner-most planet that survives the violent evolution resulting from
the mutual orbit crossing.

Of course, it is quite possible that some fraction of the observed
spin-orbit misalignment is of a primordial origin.  The asteroseismic
analysis by \citet{Huber2013}, for instance, revealed that the stellar
inclination of Kepler-56 with two transiting planets is approximately
$45^\circ$. While it is possible to dynamically change the orbital
plane of the two planets in a coherent fashion
\citep{Huber2013,Gratia2017}, it would be more likely that the
spin-orbit misalignment observed for Kepler-56 simply reflects the
initial condition.

The above two possibilities for the origin of the spin-orbit
misalignment, which we refer to as the realignment channel and the
primordial channel, are not necessarily exclusive, and thus the
observed distribution may be accounted for by their
combination \citep[see, e.g., ][]{Winn2017}.

A possible problem for the realignment channel is that the alignment
time-scale is longer than the orbit damping
\citep{Lai2012,Rogers2013,Xue2014}.  According to the conventional
equilibrium tide model for near-circular orbits\citep{SSD}, the
semi-major axis of the planet, $a$, and the spin angular velocity of
the star, $\Omega_\star$ evolve according to
\begin{eqnarray}
\label{eq:dot-a}
  \frac{da}{dt} &=& - \frac{2}{Q'_\star} \frac{m_{\rm p}}{m_\star}
  \left(\frac{R_\star}{a}\right)^5 na, \\
\label{eq:dot-omegas}
  \frac{d\Omega_\star}{dt} &=& - {\rm sign}(\Omega_\star -n)
    \frac{1}{\alpha_\star Q'_\star} \left(\frac{m_{\rm p}}{m_\star}\right)^2
  \left(\frac{R_\star}{a}\right)^3 n^2 .
\end{eqnarray}
In the above expressions, $m_{\rm p}$ and $m_\star$ are the mass of
the planet and the star, $R_\star$ is the radius of the star, $n$ is
the mean motion of the planet, $\alpha_\star$ is the inertia moment of
the star in units of $m_\star R_\star^2$, and we introduce the effective
tidal quality factor of the star:
\begin{eqnarray}
\label{eq:q-star}
  Q'_\star \equiv \frac{2Q_\star}{3k_{2\star}}
\end{eqnarray}
with $Q_\star$ and $k_{2\star}$ being the quality factor and the second
Love number of the star, respectively.

Equations (\ref{eq:dot-a}) and (\ref{eq:dot-omegas}) imply
the corresponding damping and synchronization (or alignment)
time-scales:
\begin{eqnarray}
\label{eq:tau-a}
  \tau_{a} &\equiv& \left|\frac{a}{da/dt} \right|
  \cr
  &\approx& 3.0 \times 10^{9}
  \left(\frac{Q'_\star}{10^6}\right)
  \left(\frac{m_\star}{M_\odot}\right)^{8/3}
  \left(\frac{M_{\rm J}}{m_{\rm p}}\right)
  \left(\frac{R_\odot}{R_\star}\right)^5
  \left(\frac{P_{\rm orb}}{1 {\rm day}}\right)^{13/3} {\rm yr}, \\
\label{eq:tau-sync}
  \tau_{\rm sync}
  &\equiv& \left|\frac{\Omega_\star}{d\Omega_\star/dt}\right|
  \cr
  &\approx& 1.4 \times 10^{11}
  \left(\frac{\alpha_\star}{2/5}\right)
  \left(\frac{Q'_\star}{10^6}\right)
  \left(\frac{m_\star}{M_\odot}\right)^3
  \left(\frac{M_{\rm J}}{m_{\rm p}}\right)^2
  \left(\frac{R_\odot}{R_\star}\right)^3
  \left(\frac{P_{\rm orb}}{1 {\rm day}}\right)^3
  \left(\frac{P_{\rm orb}}{P_{\rm rot}}\right) {\rm yr},
\end{eqnarray}
where $P_{\rm orb}=2\pi/n$ and $P_{\rm rot}=2\pi/\Omega_\star$
denote the orbital period of the planet and the spin rotation period
of the star, respectively.

The (re)alignment is unlikely to be completed within the age
of the universe if one assumes a typical value of the tidal quality
factor $Q'_\star (=10^5 - 10^7)$.  Moreover, the fact of $\tau_a \ll
\tau_{\rm sync}$ regardless of the value of $Q'_\star$ implies that
the realigned planet should have been fallen into the star. Thus the
realignment channel does not seem to work in the conventional
equilibrium tide model. This is why \citet{Lai2012} proposed an
alternative tidal model \citep[see also][]{Rogers2013,Xue2014}.

Since the spin-orbit alignment is usually supposed to proceed in a
roughly similar time-scale of the orbit circularization and the
spin-orbit synchronization, one may test the realignment channel
hypothesis from the distribution of the eccentricity and $P_{\rm
  rot}/P_{\rm orb}$.  In particular, the realignment channel would
imply that $P_{\rm rot} \approx P_{\rm orb}$, while no specific
correlation is expected between $P_{\rm rot}$ and $P_{\rm orb}$ in the
primordial channel.

Unfortunately, while $P_{\rm orb}$ can be measured precisely for
transiting planets, it is not always the case for $P_{\rm rot}$ of
their host star.  It is possible to estimate $P_{\rm rot}$
spectroscopically, combining the equatorial rotational velocity from
Doppler broadening and the stellar radius. The estimate, however,
depends on the assumed turbulence, and also requires the stellar
radius and inclination that are usually not well-determined.  Although
the photometric variation of the star is more directly related to
$P_{\rm rot}$, the formation and dissipation of star-spots complicate
the interpretation of the photometrically estimated rotation period
$P_{\rm rot,phot}$.

In this respect, asteroseismology provides a complementary and more
reliable estimate for the stellar rotation period $P_{\rm rot,
  astero}$. Furthermore, since asteroseismology fits both $P_{\rm rot,
  astero}$ and the stellar inclination $\is$ \citep{Toutain1993,
  Gizon2003, Huber2013,
  Chaplin2013,Benomar2014b,Campante2016,Kamiaka2018a}, the spin-orbit
misalignment and synchronization can be examined simultaneously. Thus
asteroseismology is a unique methodology to probe the spin-orbit
architecture of the transiting planetary systems, and also to test
empirically the degree of the star-planet tidal interaction in a
model-independent fashion.

The analysis of the $P_{\rm rot, phot}/P_{\rm orb}$ has been performed
for {\it Kepler} eclipsing binaries (EBs) by \citet{Lurie2017}. They
measured $P_{\rm rot, phot}$ for 816 EBs from their star-spot
modulation, and found that 79\% of EBs with $P_{\rm orb}<10$~days are
synchronized. They also noted that the fraction of super-synchronous
($P_{\rm orb} > P_{\rm rot}$) EBs significantly increases for $P_{\rm
  orb}>10$~days. The tidal interaction between the host star and
planets in exoplanetary systems should be much weaker than that
between stars in EBs. Nevertheless we found a similar tendency for
three {\it Kepler} transiting planetary systems, as will be
shown below in detail.

The rest of the paper is organized as follows. Section
  \ref{sec:Prot} critically compares the stellar rotation periods
  estimated from photometric variation and asteroseismology. We find
  that $P_{\rm rot, phot}$ is somewhat sensitive to the detail of the
  underlying assumptions and needs to be interpreted with caution.
  Section \ref{sec:implication} describes two major implications from
  the simultaneous measurements of $P_{\rm rot, phot}$ and $P_{\rm
    rot, astero}$; asteroseismic constraints on stellar inclination
  and frequency splitting, and a possible signature of spin-orbit
  resonance.  Finally the summary of the paper is presented in Section
  \ref{sec:conclusion}.  

\section{Stellar rotation period from photometric variation and
   asteroseismology  \label{sec:Prot}}

\subsection{Our sample of {\it Kepler} transiting planetary systems
   for asteroseismology  \label{subsec:sample}}

In many cases, $P_{\rm rot, phot}$ derived from photometric variation
is more {\it precise} than $P_{\rm rot, astero}$ from
asteroseismology. It does not necessarily imply, however, that $P_{\rm
  rot, phot}$ is more {\it accurate} than $P_{\rm rot, astero}$. The
present analysis adopts a sample of 33 stars with transiting planets
from {\it Kepler} data, which are analyzed with asteroseismology by
\citet{Kamiaka2018a}.

We focus on systems whose stellar rotation periods are relatively well
measured from asteroseismology.  Specifically we select 19 systems for
which $v_\star \sin i_\star$ from asteroseismology is inconsistent
with $0$ within 5$\sigma$ (Table \ref{tab:stellar_property}).

The stellar rotation of those systems is fast enough to securely
measure the rotation period from their power spectra. For comparison
and reference, we also consider 48 stars without known planets, but
with reliable $v\sin{\is}$ measurement, out of 61 {\it Kepler} stars
analyzed in \citet{Kamiaka2018a}.  Among these $19+48=67$ stars, 30
objects are also analyzed independently by \citet{Benomar2018}.  We
find that the two independent estimates of $P_{\rm rot,astero}$ for 26
among the 30 stars agrees within $1\sigma$, and that the remaining 4
are consistent within $2\sigma$, suggesting that the
asteroseismic result is almost free from details of the individual
analysis.

\subsection{The Lomb-Scargle periodogram
  for photometric stellar rotation periods \label{subsec:LS}}

We compute the Lomb-Scargle (LS) periodogram of the 19 planet-host stars from
their long cadence PDCSAP lightcurves provided on the KASOC website
(\url{http://kasoc.phys.au.dk}). Quarters are first concatenated by
fitting the fourth-order polynomials on each quarter and extrapolating
the time to the initial time of the subsequent quarter.  This
efficiently removes {\it jumps} due to the change of CCD when {\it
  Kepler} rotates. while preserving temporal gaps between
quarters. Additionally, a smooth curve (a box-car smoothing of 50 days
width) is removed from the concatenated lightcurve in order to
effectively filter out variabilities longer than $\approx 50$ days.

Effects of transits on photometric variation are minimized by trimming
the lightcurve. To find the best trimming threshold, we visually
inspect each lightcurve on a trial-and-error basis. We
also verify that the signal from the transits is effectively removed
from the low frequency part of the LS periodogram. Note that the LS
periodogram is computed using an oversampling factor of four.

A low-frequency peak of the LS periodogram is interpreted as the
surface rotation rate of the stars, due to surface structures
co-rotating with the stellar surface.  To minimize noise fluctuations,
the peak position is extracted from the LS periodogram smoothed over a
box-car window of width $0.1\,\mu$Hz. This value corresponds to the
typical width of the observed peak and might be due to the finite
lifetime of surface star-spots and/or the latitudinal
differential rotation. The peak extraction is performed over the range
$0.2 - 3.0\,\mu$Hz, corresponding to periods between 3.8 and 60 days.

The uncertainty on the peak position is estimated from its
full-width-at-half-maximum of power in the {\it frequency}, instead of
time, domain.  We compute the corresponding frequency region in a
linearly equally bin in the frequency, and convert it in the time
domain, which is indicated as blue-shaded regions in Figure
\ref{fig:periodogram}.  This works nicely for the 13 {\it reliable}
stars with a clear peak in the periodogram, but the resulting
error-bars in $P_{\rm rot, phot}$ are fairly uncertain, and should not
be trusted for the other six stars (labeled {\it bimodal} or {\it
  uncertain} in what follows).

\subsection{Comparison with previous photometric variation
  \label{subsec:comparison}}

\begin{figure}[ht]
  \centering
  \includegraphics[width=11cm]{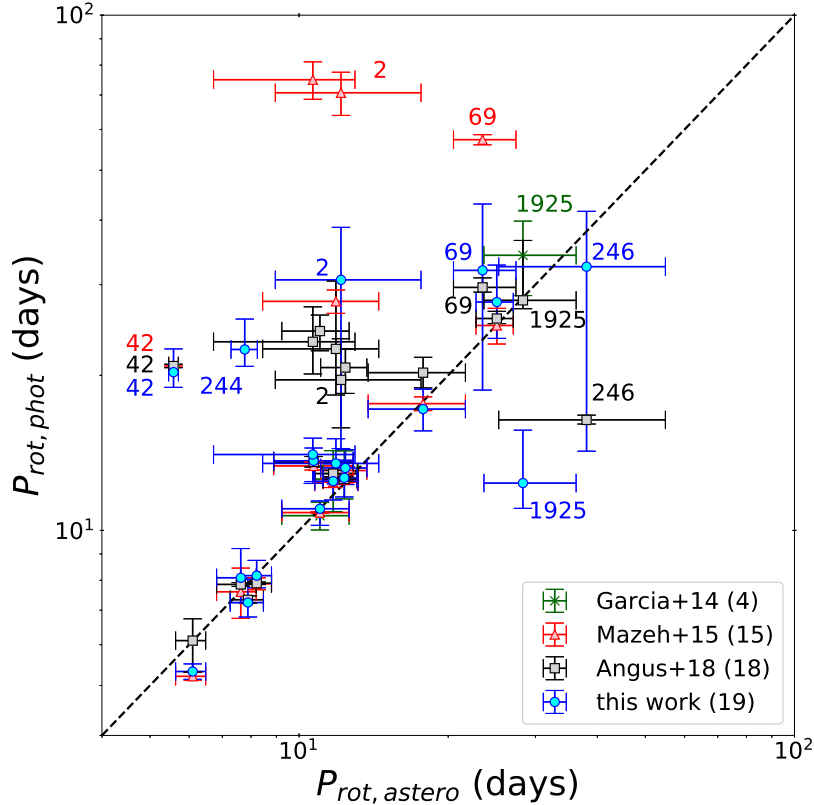}
\caption{Photometric rotation periods $P_{\rm rot, phot}$ of the 19
  planet-host stars against their asteroseismic rotation periods
  $P_{\rm rot, astero}$.  The values of $P_{\rm rot, phot}$ are based
  on four independent papers as indicated in the legend. The number in
  the parenthesis indicates the number of stars plotted here that are
  overlapped in the paper and this work. We mark 6 stars, whose
  $P_{\rm rot, phot}$ derived from the LS periodogram is unreliable
  ({\it bimodal} or {\it uncertain}), by their KOI IDs (see Table
  \ref{tab:stellar_property}).}
\label{fig:period_comparison}
\end{figure}

Figure \ref{fig:period_comparison} plots $P_{\rm rot,phot}$ for the 19
planet-host stars against our value of $P_{\rm rot,astero}$.  We plot
the values of $P_{\rm rot,phot}$ for 4 stars by \citet{Garcia2014}
with the Morlet wavelet in green, 15 stars by \citet{Mazeh2015b} with
the auto-correlation function in red, and 18 stars by
\citet{Angus2018} with Gaussian process in gray.  Our own measurement
of $P_{\rm rot,phot}$ using the LS periodogram is plotted in blue.

Clearly, measured values of $P_{\rm rot,phot}$ published in literature
are rather different, indicating that the measurements of $P_{\rm rot,
  phot}$ are somewhat dependent on the detailed methods of identifying
the photometric variation. This is why discrepant values for the same
systems are exhibited in some cases.  In particular, we note that for
$P_{\rm rot, astero}\approx 10 - 20$ days, the estimates by
\citet{Angus2018} are larger by a factor $\approx 2$ (gray squares)
relative to ours (blue circles). We individually examine the the LS
periodogram of the 19 systems, and find that their estimates do not
correspond to the highest peaks for most of the above cases.

As \cite{Angus2018} clearly mentioned, Gaussian Processes (GP) are
prone to over-fitting and require a lot of care when setting the
hyper-parameters and hyper-priors. Actually, our examination of the
low frequency power spectrum suggests that the GP method picks up a
time-scale consistent with that of the convective turnover expected
for Sun-like stars \citep[see e.g.][]{Landin2010}, rather than the
stellar rotation period. Therefore, it is likely that the GP method is
difficult to clearly distinguish the granulation noise (in the power
spectrum it shows up as a pink noise, often referred to as the
Harvey-like profile) from the signal corresponding to the stellar surface
rotation.

\begin{figure}[ht]
  \centering
\includegraphics[width=12cm]{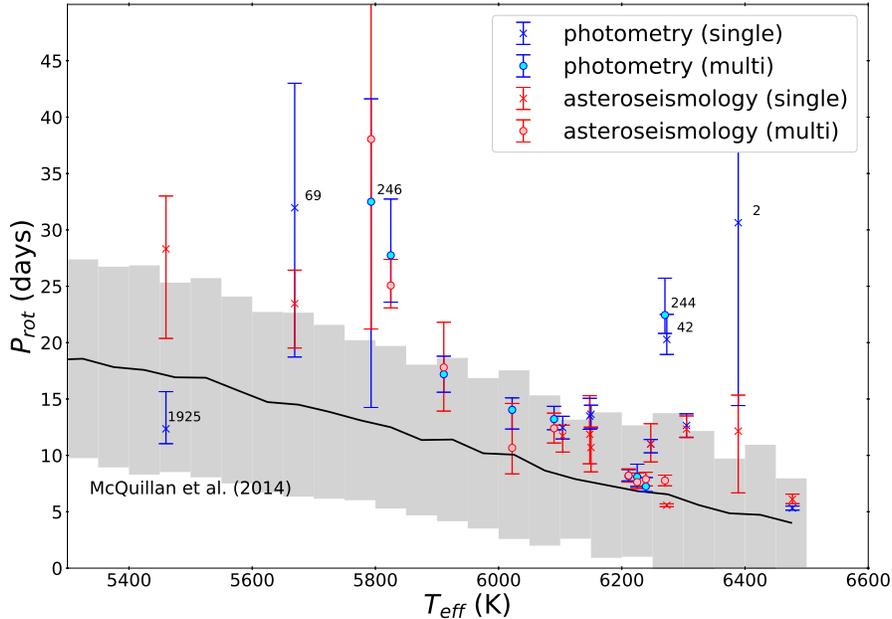}
\caption{Rotation periods of the 19 stars against their effective
  temperature. Blue and red symbols correspond to $P_{\rm rot, phot}$
  and $P_{\rm rot, astero}$ with crosses and circles indicating single
  and multiple planet systems, respectively.  The mean and its
  1$\sigma$ uncertainty regions for the photometrically derived
  rotation period \citep{McQuillan2014} are plotted as the thick black
  line and the gray area.  We mark 6 stars, whose $P_{\rm rot, phot}$
  derived from the LS periodogram is unreliable by their KOI IDs.}
\label{fig:Teff-Prot}
\end{figure}

Both our asteroseismic and photometric estimates are largely
consistent with the result of \citet{Mazeh2015b} plotted in red
triangles, but there are three stars for which their auto-correlation
method gives rotational periods of more than $\sim 60$ days.  This
could be due to our box-car smoothing of 50 days (see \S
\ref{subsec:LS}), but is statistically unexpected for a Sun-like star
in the main sequence phase \citep[see][and our Fig \ref{fig:Teff-Prot}
  below]{McQuillan2014}. Therefore we suspect that they correspond to
harmonics of the true rotation period, perhaps more visible in the
auto-correlation function adopted by \citet{Mazeh2015b} rather than in
the LS periodogram.

Figure \ref{fig:Teff-Prot} shows $P_{\rm rot, astero}$ (red circles)
estimated by asteroseismology and $P_{\rm rot, phot}$ (blue circles)
estimated by LS periodogram for those 19 planet-host stars against the
stellar effective temperature $T_{\rm eff}$.  For comparison, the mean
and 1$\sigma$ region of $P_{\rm rot}$ -- $T_{\rm eff}$ from
photometric variation analysis of $\simeq 34,000$ Kepler stars
\citep{McQuillan2014} are plotted as the thick black line and gray
area, respectively.

Clearly both $P_{\rm rot, astero}$ and $P_{\rm rot, phot}$ measured by
us for the 19 stars are systematically longer than the average of {\it
  Kepler} stars by \citet{McQuillan2014}.  As indicated by Figure
\ref{fig:period_comparison}, this tendency becomes even stronger if we
plot $P_{\rm rot, phot}$ by \citet{Mazeh2015b} and \citet{Angus2018}.
We also made sure that 48 planet-less stars with secure rotational
period measurements from \citet{Kamiaka2018a} exhibit the same trend,
implying that the systematic tendency is not related to the effect of
the accompanying planet.

The reason for this difference is unclear, but we suspect that this
results from (unknown) factors affecting the detectability of
solar-like pulsations. For example, magnetic activity is known to damp
solar pulsations so that they show reduced amplitudes
\cite[e.g.][]{Benomar2012}.  The statistical distribution derived by
\citet{McQuillan2014}, however, is still consistent at $2\sigma$ with
our estimates, and thus the apparent discrepancy may be simply due to
the limited size of our sample.

\subsection{Validation of reliability of the stellar rotation periods
  \label{subsec:validation}}

\begin{figure}[ht]
  \centering
\includegraphics[width=10cm]{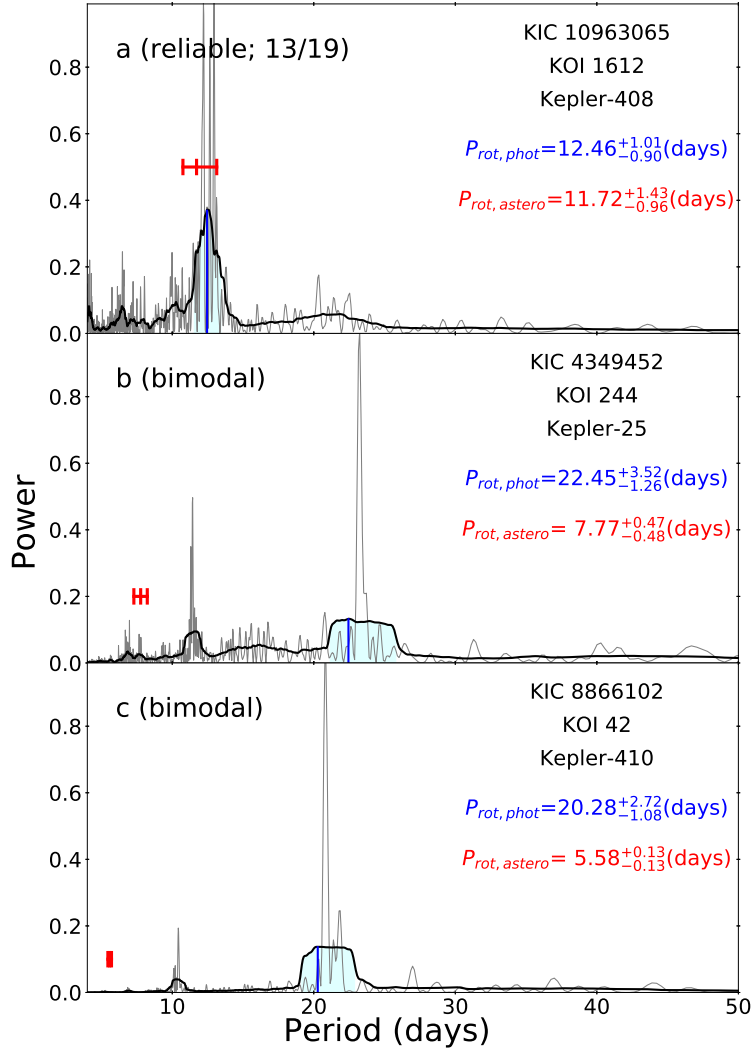}
\caption{Examples of the LS periodogram for our sample.  The thick
  black line indicates the boxcar-smoothed result (over 0.1 $\mu$Hz)
  of the original LS periodogram (thin gray curves). The original
  periodogram is normalized so that the maximum power of each system
  is unity.  The period corresponding to the maximum power of the
  smoothed LS curve is marked by the vertical blue line, and the
  associated range of its full-width-at-half-maximum is plotted as
  blue-shaded areas.  We also show the mean and its $1\sigma$
  confidence interval of the asteroseismic rotation period by the
  horizontal red bar.  Panel {\it a}. Example of {\it reliable}
  systems with a clear signature of the photometric rotation.  Panels
  {\it b} and {\it c}. two {\it bimodal} systems (Kepler-25 and 410)
  that exhibit clear double peaks, neither of which match the
  asteroseismic rotation period.}
\label{fig:periodogram}
\end{figure}

Our LS periodogram analysis returns unusually large uncertainties for
four KOIs (KOI 2, 69, 246, and 1925), and discrepant results compared
to seismology for two KOIs (KOI 42 and 244), which are labeled in
Figures \ref{fig:period_comparison} and \ref{fig:Teff-Prot}.  We
carefully examine their LS periodogram, and consider the origin of these
discrepancies as described in what follows.

Figure \ref{fig:periodogram}a shows the LS periodogram for KOI-1612
(Kepler-408) whose highest peak (blue area) is consistent with the
period estimated from asteroseismology (red bars); 13 out of the 19
systems belong to this case, and will be referred to as {\it
  reliable}. We note also that Kepler-408b is the smallest planet ever
discovered to be in a significantly misaligned orbit
\citep{Kamiaka2019}.

Figure \ref{fig:periodogram}b and c plots the LS periodogram for the
two stars classified as {\it bimodal}, , KOI-244 (Kepler-25) and
KOI-42 (Kepler-410), which exhibit a discrepancy between seismology
and the LS periodogram analysis.  We note that they have two clear
peaks in the LS periodogram, neither of which agrees with the seismic
rotation period.

We do not yet understand the origin of this bimodality nor
discrepancy.  It may indicate that the transit signal is not
completely removed during the lightcurve preparation, and that the
residual contaminates the periodogram. It seems more likely, however,
that the peaks are related to some harmonics of the true rotation
period, while the true period itself is obscured for some unknown
reason. Indeed $P_{\rm rot, phot}$ corresponding to the highest peak
in the periodogram are $\approx 3 P_{\rm rot, astero}$ and $\approx 4
P_{\rm rot, astero}$ for KOI-244 and KOI-42, respectively.

The remaining four systems, referred to as {\it uncertain}, KOI-2,
246, 69 and 1925, do not show any clear peak in the LS periodogram. We
cannot estimate the rotation period of those stars due to the large
uncertainty.  This may be partly because the star is significantly
inclined with respect to the line-of-sight, or partly because the star
has a weak magnetic activity level.  For cool stars that are supposed
to exhibit detectable star-spot activity, therefore, such transiting
planetary systems with no clear peak in the periodogram may be good
candidates for oblique systems.  This will be discussed further in
subsection \ref{subsec:is} below.

\section{Implications  \label{sec:implication}}

\subsection{Constraints on stellar inclination and frequency
  splitting from asteroseismology
  \label{subsec:is}}

  In order to see if the four {\it uncertain} stars indeed correspond to
oblique (low inclination) systems, we compute constraints on their
stellar inclination and the rotational splitting $\delta\nu_\star$
from asteroseismic analysis. Further details of the analysis are
described in \citet{Kamiaka2018a,Kamiaka2019}.

\begin{figure}[ht]
\centering
\includegraphics[width=11cm]{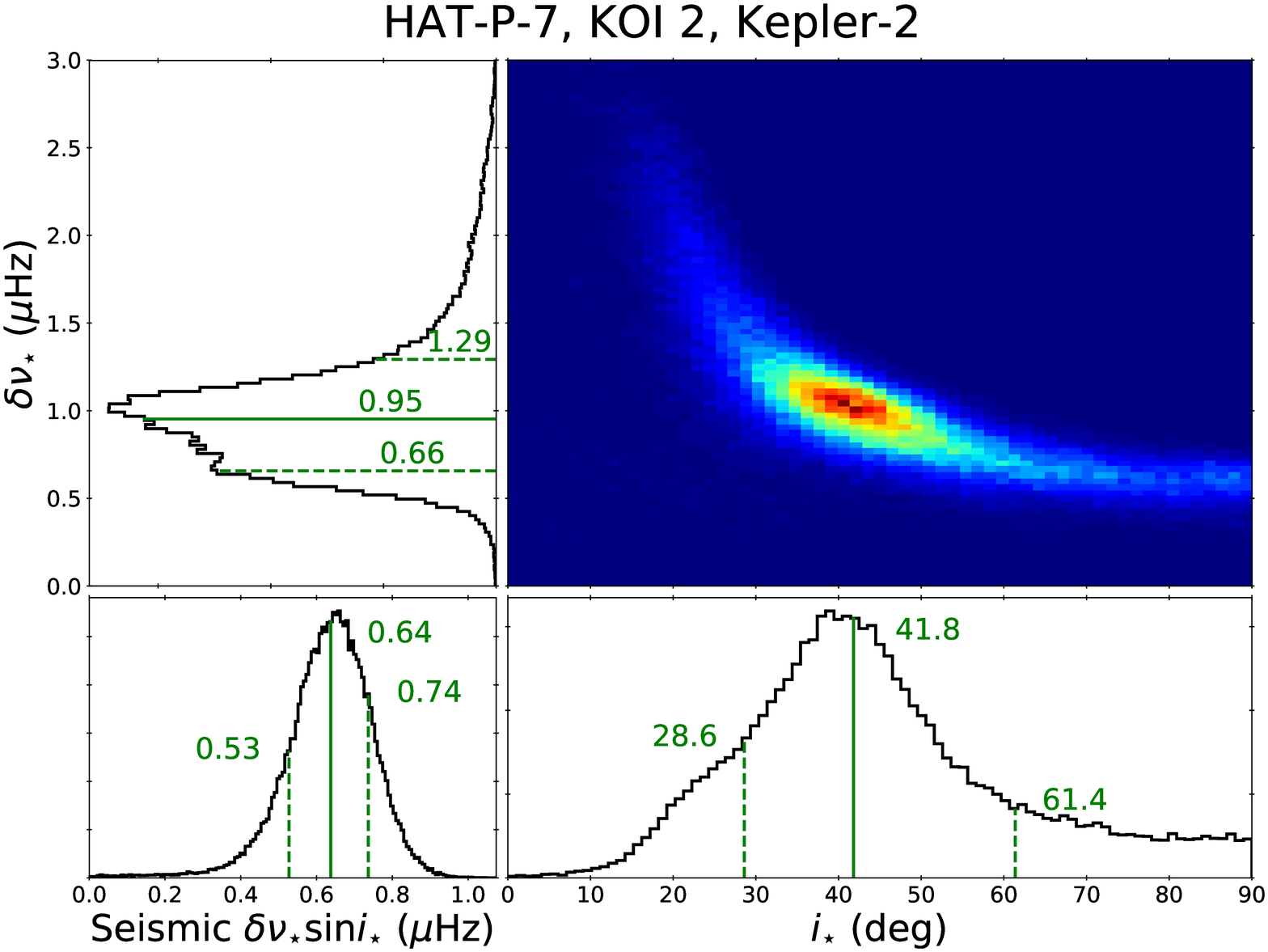}
  \centering

\includegraphics[width=11cm]{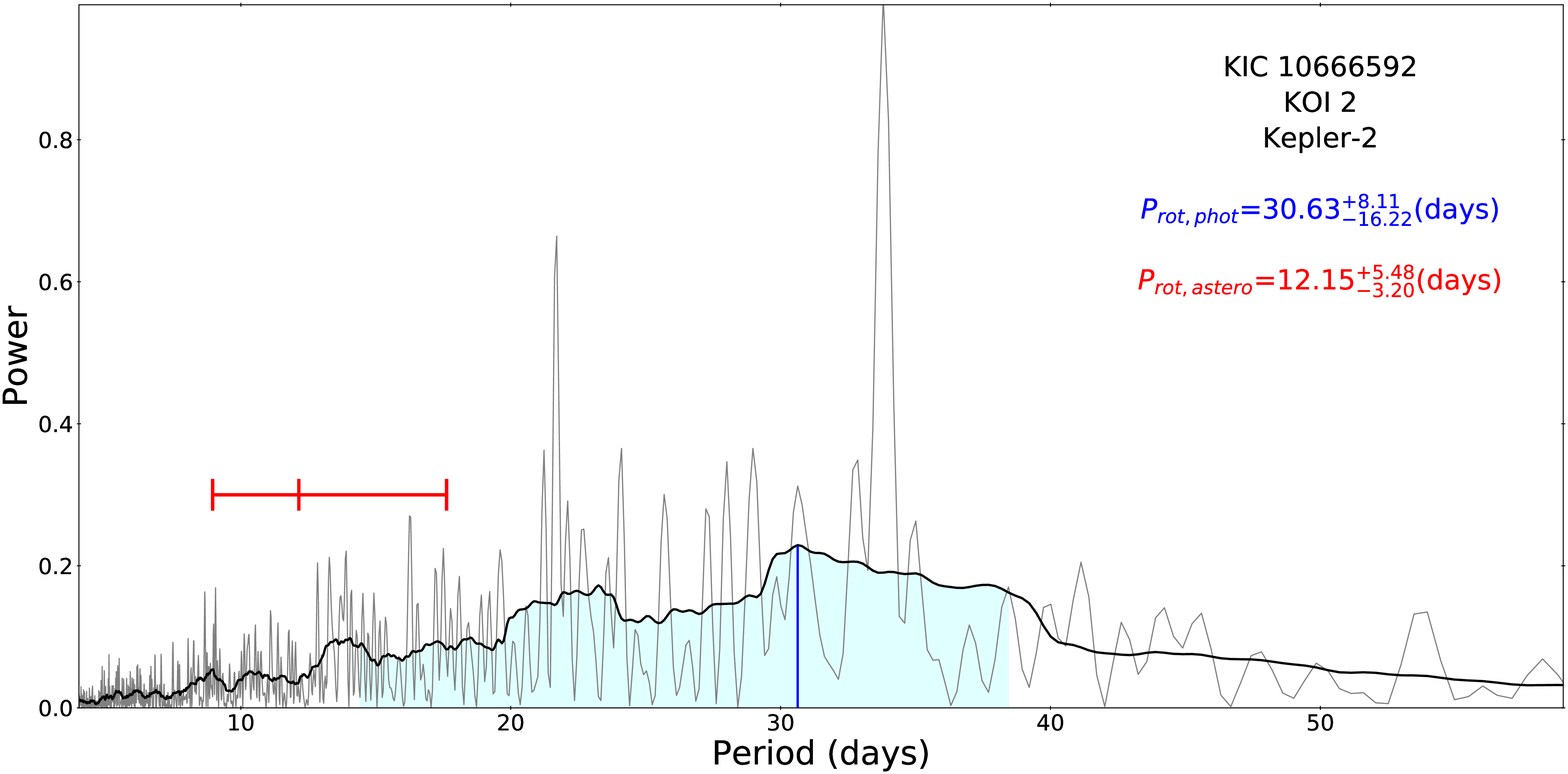}
\caption{Constraints on the stellar inclination $\is$ and frequency
  splitting $\spl$ of KOI-2 (Kepler-2, HAT-P-7) from asteroseismic analysis.
  We plot the posterior probability density (PPD) on $\is$ -- $\spl$
  plane, marginalized over the other parameters.  The one-dimensional
  marginalized densities are also shown to the left and below the
  axes.  The bottom--left panel is the PPD of $\spl\sin{\is}$.
}
  \label{fig:kepler2}
\end{figure}

\begin{figure}[ht]
\centering
\includegraphics[width=11cm]{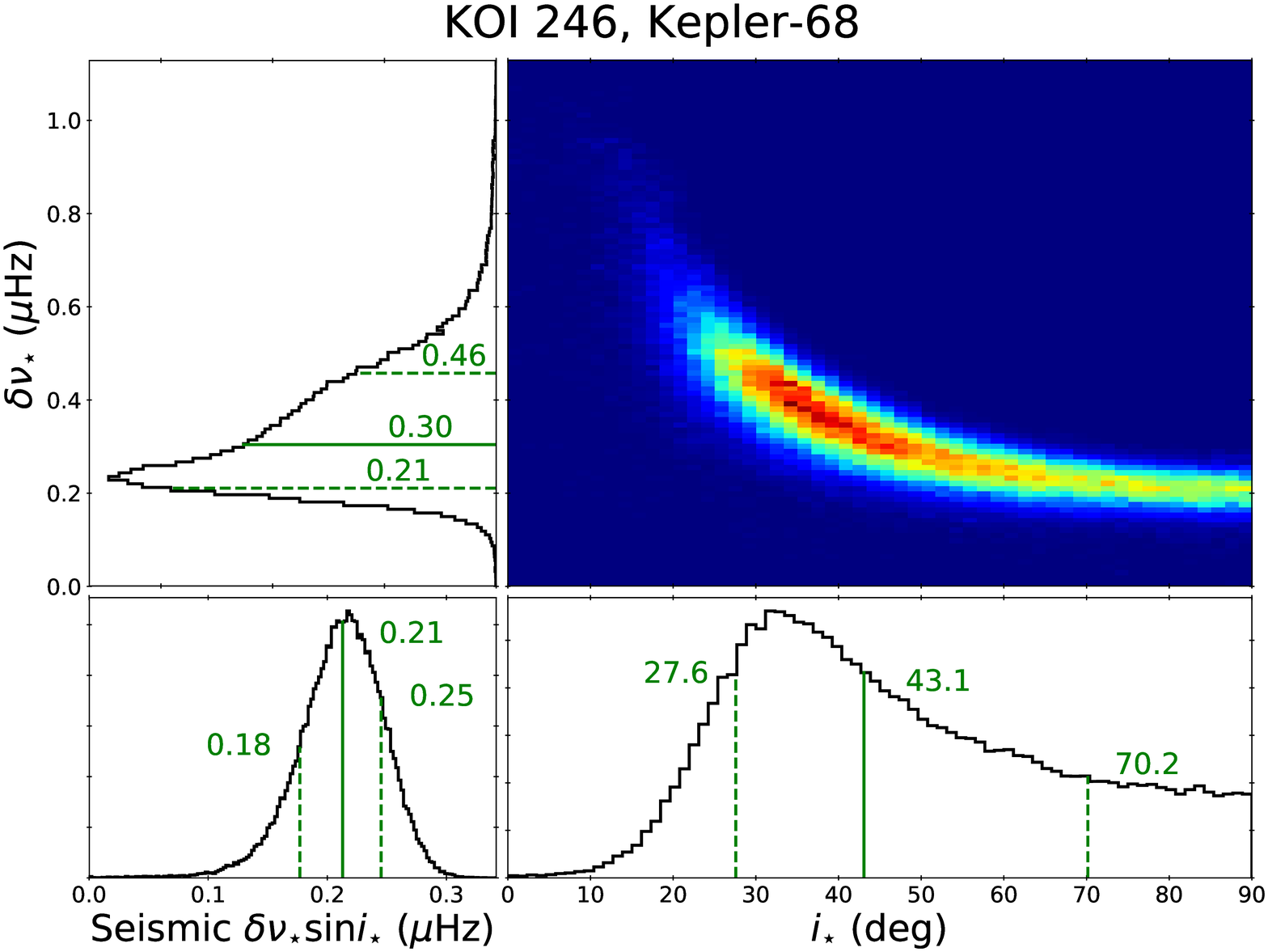}
  \centering

\includegraphics[width=11cm]{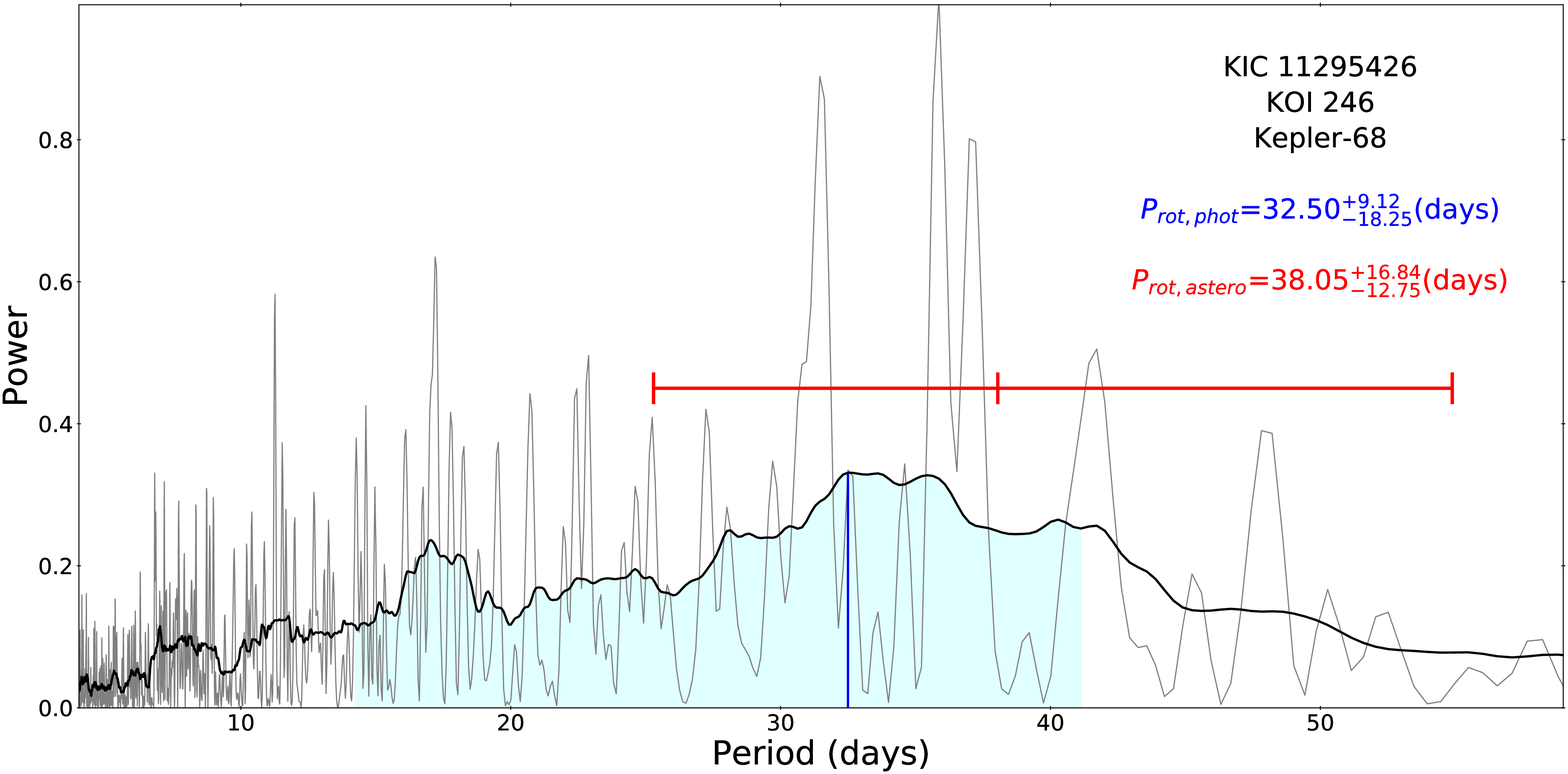}
\caption{Same as Figure \ref{fig:kepler2} but for
  KOI-246 (Kepler-68).  }
  \label{fig:kepler68}
\end{figure}

\begin{figure}[ht]
\centering
\includegraphics[width=11cm]{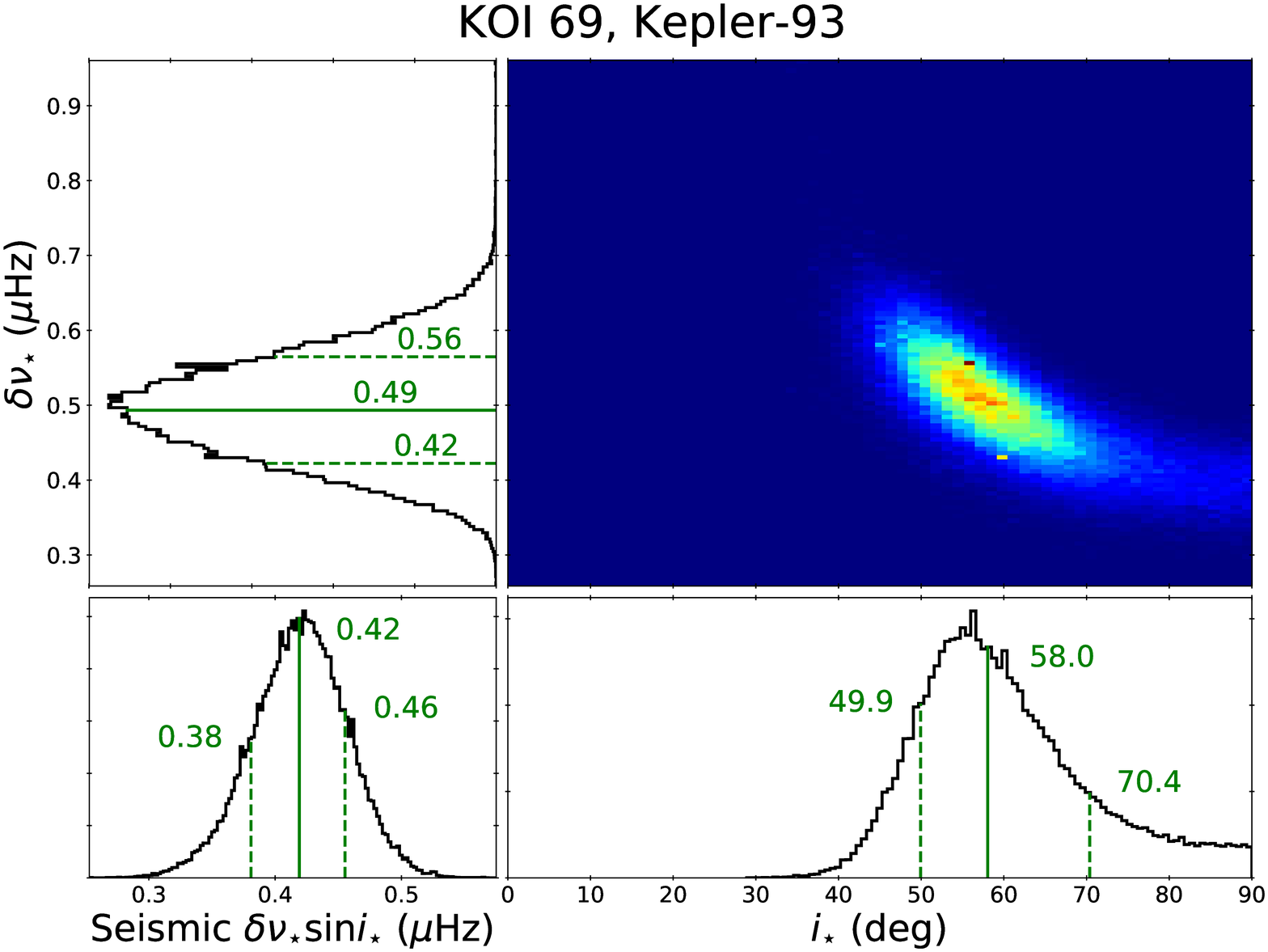}
  \centering

\includegraphics[width=11cm]{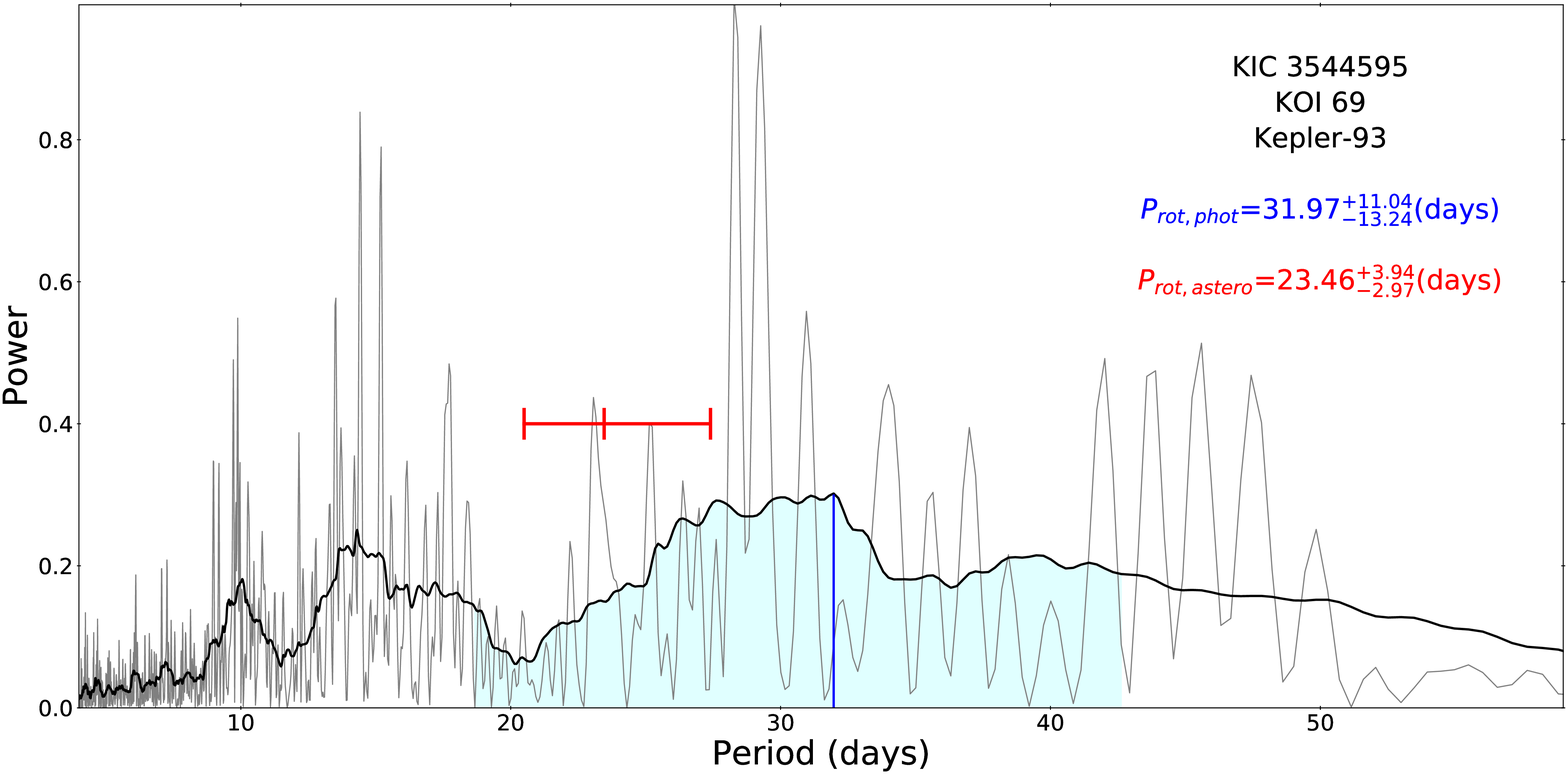}
\caption{Same as Figure \ref{fig:kepler2} but for
  KOI-69 (Kepler-93).  }
  \label{fig:kepler93}
\end{figure}

\begin{figure}[ht]
\centering
\includegraphics[width=11cm]{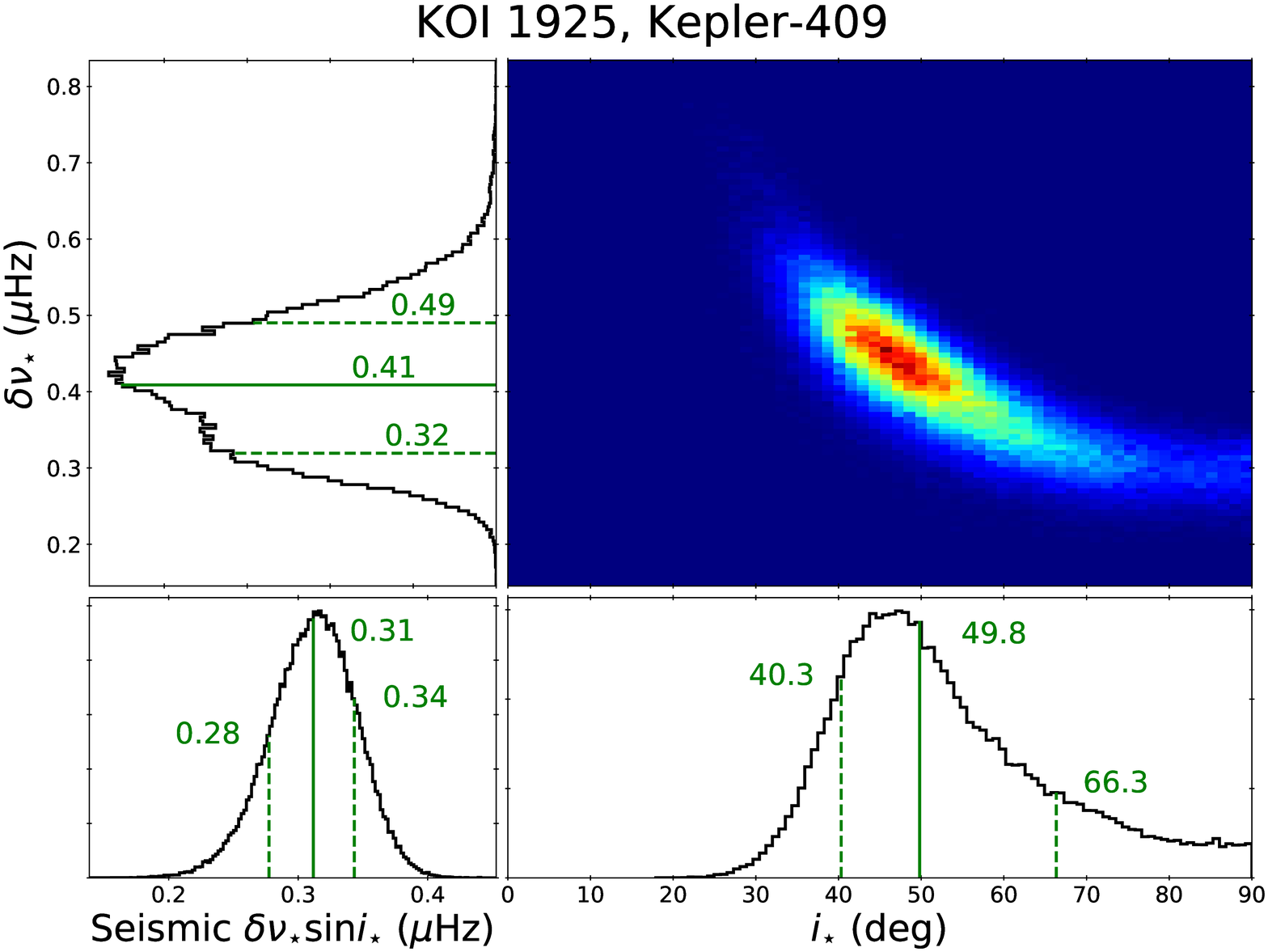}

  \centering
\includegraphics[width=11cm]{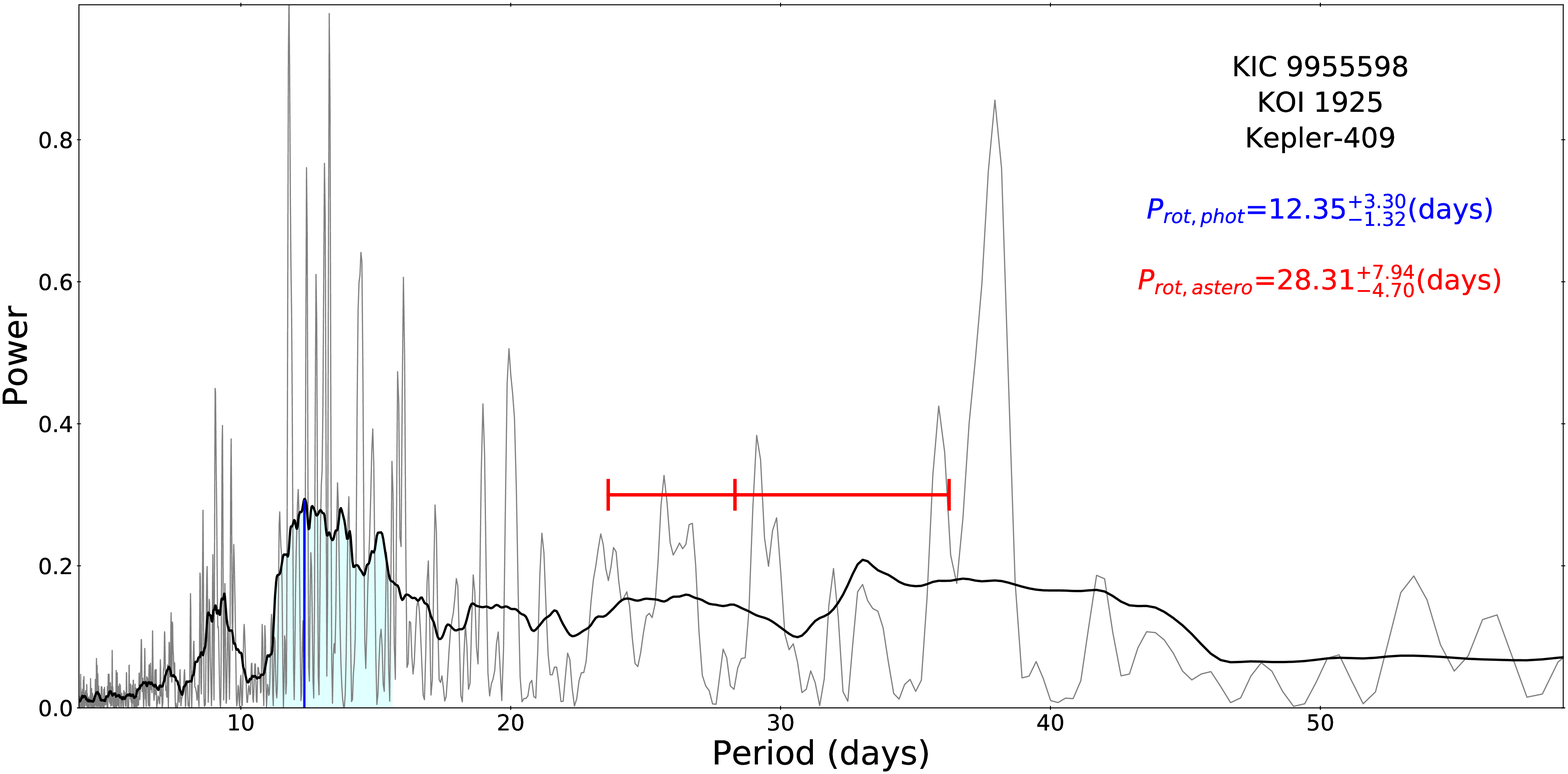}
\caption{Same as Figure \ref{fig:kepler2} but for KOI-1925
  (Kepler-409). }
\label{fig:kepler409}
\end{figure}

The top-right panels in Figures \ref{fig:kepler2} to
  \ref{fig:kepler409} show the posterior probability density (PPD) on
  $\is$ -- $\spl$ plane, marginalized over the other model parameters.
  The corresponding one-dimensional marginalized densities of $\spl$
  and $\is$ are plotted to the left and below the axes, respectively.
  Finally the bottom--left panel is the marginalized PPD of
  $\spl\sin{\is}$, which may be estimated independently from the
  spectroscopic analysis of the line profiles. The solid and dashed
  lines in the one-dimensional marginalized PPD indicate the median
  and the associated 68\% credible ranges, respectively.  

If neither latitudinal nor radial differential rotation is present,
the stellar rotation period is simply related to the rotational
splitting $\delta\nu_\star$ estimated from asteroseismology as
\begin{equation}
  P_{\rm rot} \approx 11.6
  \left(\frac{\mu {\rm Hz}}{\delta\nu_\star} \right){\rm days} .
\end{equation}

  As Figures \ref{fig:kepler2} to \ref{fig:kepler409} indicate, $\is$
and $\spl$ ($=1/P_{\rm rot, astero}$) are strongly correlated in
general.  On the other hand, the asteroseismic analysis is known to be
able to identify the value of $\spl\sin\is$ in a robust manner.
\citet{Kamiaka2019} have presented the most comprehensive
discussion on the joint analysis of asteroseismology and photometric
variation for Kepler-408, one of the {\it reliable} stars in the
present sample.

  Unfortunately $P_{\rm rot, phot}$ for the {\it uncertain} stars are
not reliable, and we cannot break the degeneracy precisely.
Nevertheless, it is clear that they have systematically lower
inclinations around $40^\circ$ than the {\it reliable} stars from
asteroseismology alone, except Kepler-408 (see Table
\ref{tab:stellar_property}).  This supports, at least qualitatively,
our interpretation why they do not show any detectable periodicity in
their photometric lightcurves.

In this context, we emphasize that a large {\it projected} spin-orbit
misalignment of Kepler-2 (HAT-P-7) has already been discovered by
\citet{Winn2009}.  Moreover, \citet{Benomar2014} attempted for the
first time to recover the full spin-orbit angle, instead of its
projected value $\lambda$, through the joint analysis of the RM effect
and asteroseismology. They considered two systems, HAT-P-7 (Kepler-2,
Figure \ref{fig:kepler2}) and Kepler-25 (Figure
\ref{fig:periodogram}b), which are classified here as {\it uncertain} and
    {\it bimodal}, respectively.

      As in the case of Kepler-56 \citep{Huber2013}, the determination of
the stellar inclination angles of Kepler-68 and Kepler-25 are of great
value because they host more than one transiting planets.  Kepler-68
has three planets, including two inner rocky planets ($R_{\rm
  p}=2.4R_{\oplus}, 1.0R_{\oplus}$) in compact, and possibly
eccentric, orbits ($P_{\rm orb}=5.4$ days, $9.6$ days); see Table
\ref{tab:planet_multi}. Our analysis indicates that $\is =
43.1^{+27.1}_{-15.1}$ and $80.6^{+6.6}_{-9.2}$ degrees for Kepler-68
and Kepler-25, respectively. While Kepler-68 could be another case for
the strongly inclined multi-planetary system like Kepler-56, it is
still consistent with $\is=90$ degrees as shown in Figure
\ref{fig:kepler68}.
    
Kepler-93 has a close-in rocky planet ($R_{\rm p}=1.6R_{\oplus}$,
$P_{\rm orb}=4.7$ days) and a massive planet in a distant orbit
($P_{\rm orb} > 1460$ days).  Kepler-409 has an Earth-sized planet
($R_{\rm p}=1.2 R_\oplus$) in a 69-day orbit.

Because the measurement of the projected spin-orbit angle $\lambda$
for such small planets is practically impossible at this point, the
above three systems may be new interesting candidates for obliquity
studies based on asteroseismology, in particular Kepler-68 among
others\citep{Kamiaka2019}.

 It is also possible to constrain the value of $\spl\sin{\is}$
  combining the stellar radius from the {\it Kepler} photometry and
  the sky-projected rotation velocity from the observed spectral line
  broadening.  Adopting the spectroscopic measurement by
  \citet{Petigura:2017aa} and \citet{Johnson:2017aa}, we find
  $\spl\sin{\is} = 0.60 {\pm}0.12$ $\mu$Hz for HAT-P-7 (Kepler-2),
  which is in good agreement with our asteroseismic result (Figure
  \ref{fig:kepler2}). HAT-P-7 is a well-known system with a large
  projected spin-orbit angle $\lambda$, and it is reasonable that the
  stellar spin is also significantly inclined towards us
  \citep{Benomar2014}.  On the contrary, the spectroscopic estimate of
  $\spl\sin{\is} = 0.62{\pm}0.26$ $\mu$Hz for Kepler-409 is larger
  than our asteroseismic estimate, although barely consistent within
  1-2 $\sigma$ (Figure \ref{fig:kepler409}). Thus Kepler-409 may be a
  well-aligned system.  For the other two stars, Kepler-93 and 68,
  their line broadening widths are consistent with zero within an
  error of 1km/s \citep{Petigura:2017aa,Johnson:2017aa}, supporting
  our interpretation that they are significantly misaligned systems.
  Therefore those {\it uncertain} planet-host stars that exhibit no
  clear photometric variation deserve further detailed studies as good
  candidates of misaligned systems, in particular if their line
  broadening widths are unusually small.

\subsection{Possible signature of (quasi-)spin-orbit resonance
  \label{subsec:resonance}}

Given the comparison of the different estimates of $P_{\rm rot, phot}$
described above, we decided to use our own results (blue circles in
Figure \ref{fig:period_comparison}) and $P_{\rm rot, astero}$
\citep{Kamiaka2018a} as the two independent proxies for the true
rotation period in this subsection.  Because we inspected the LS
periodogram of the 19 systems individually and homogeneously, our
estimate of $P_{\rm rot, phot}$ is more robust and reliable than those
presented in the previous literature (Figure
\ref{fig:period_comparison}).

Before proceeding, we would like to stress here that strictly
speaking, neither $P_{\rm rot, astero}$ nor $P_{\rm rot, phot}$ may
represent the true rotation period of the star $P_{\rm rot, true}$.
The surface differential rotation would lead to $P_{\rm rot, phot} >
P_{\rm rot, true}$ for most stars in which the high-latitude surface
rotates more slowly than the equator. Multiple formation/dissipation
of star-spots may result in $P_{\rm rot, phot}/P_{\rm rot, true}$
significantly different from unity. It may be also the case for
$P_{\rm rot, astero}$, which mainly probes the stellar internal
rotation using its effect on stellar surface oscillations.

Taking account of a possibility that neither $P_{\rm rot,phot}$ nor
$P_{\rm rot, astero}$ does not necessarily represent $P_{\rm rot,
  true}$, we select 13 stars (out the 19 stars excluding the two
bimodal and four uncertain systems) satisfying $0.7<P_{\rm
  rot,phot}/P_{\rm rot, astero}<1.3$; see Tables
\ref{tab:planet_single} and \ref{tab:planet_multi}.  Thus their
$P_{\rm rot, astero}$ and $P_{\rm rot, photo}$ may be regarded as a
reasonably good proxy for $P_{\rm rot, true}$, while their
quantitative difference needs to be kept in mind in understanding the
result presented below.

\begin{figure}[ht]
  \centering
\includegraphics[width=12cm]{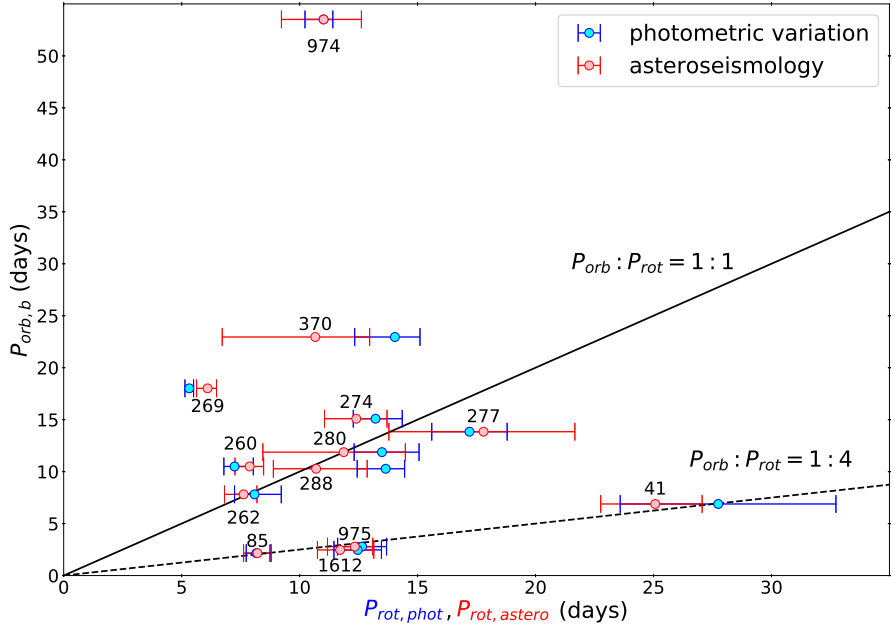}
\caption{Orbital periods of the innermost planets $P_{\rm orb, b}$
  plotted against the photometric and seismic stellar rotation
  periods, $P_{\rm rot,phot}$ (blue) and $P_{\rm rot, astero}$ (red).
  The number labeling the symbols indicates the KOI ID.
  Just for reference, $P_{\rm orb,b}/P_{\rm rot}$=1, and 1/4 are
  plotted in solid and dashed lines.  }
\label{fig:Porb-Prot}
\end{figure}

Figure \ref{fig:Porb-Prot} plots $P_{\rm orb, b}$ against $P_{\rm
  rot,phot}$ (blue) and $P_{\rm rot, astero}$ (red) for the 13
reliable systems. We identify a weak clustering around $P_{\rm orb,
  b}/P_{\rm rot}=1$ and $1/4$, especially for $P_{\rm rot, astero}$
that we assume to be more accurate and robust than $P_{\rm rot, phot}$.

\begin{figure}[ht]
  \centering
\includegraphics[width=14cm]{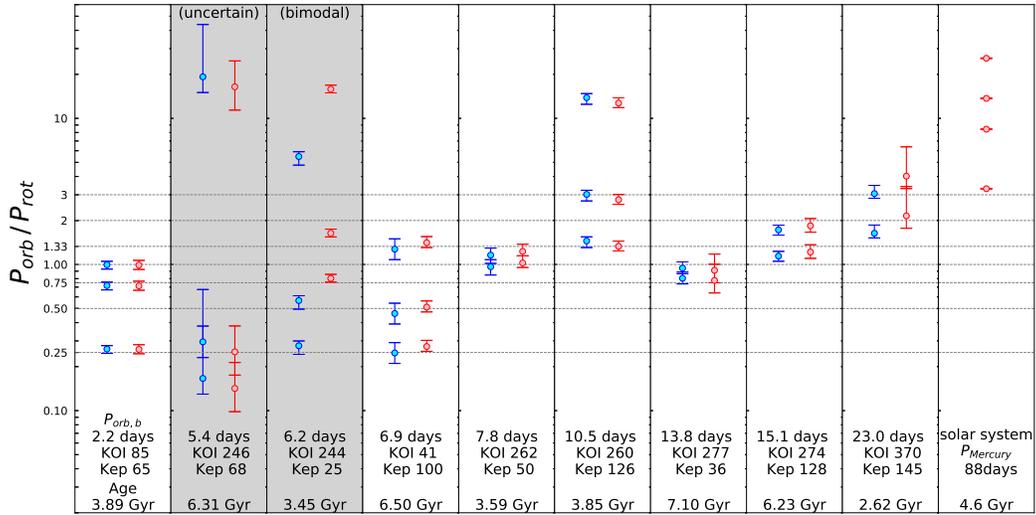}
\caption{ $P_{\rm orb}/P_{\rm rot, phot}$ (blue symbols) and $P_{\rm
    orb}/P_{\rm rot, astero}$ (red symbols) for multi-planetary
  systems. Systems without reliable $P_{\rm rot, phot}$ measurement
  are gray-shaded. Just for comparison, we plot the Solar system as
  well.}
\label{fig:Pratio-multi}
\end{figure}

\begin{figure}[ht]
  \centering
\includegraphics[width=14cm]{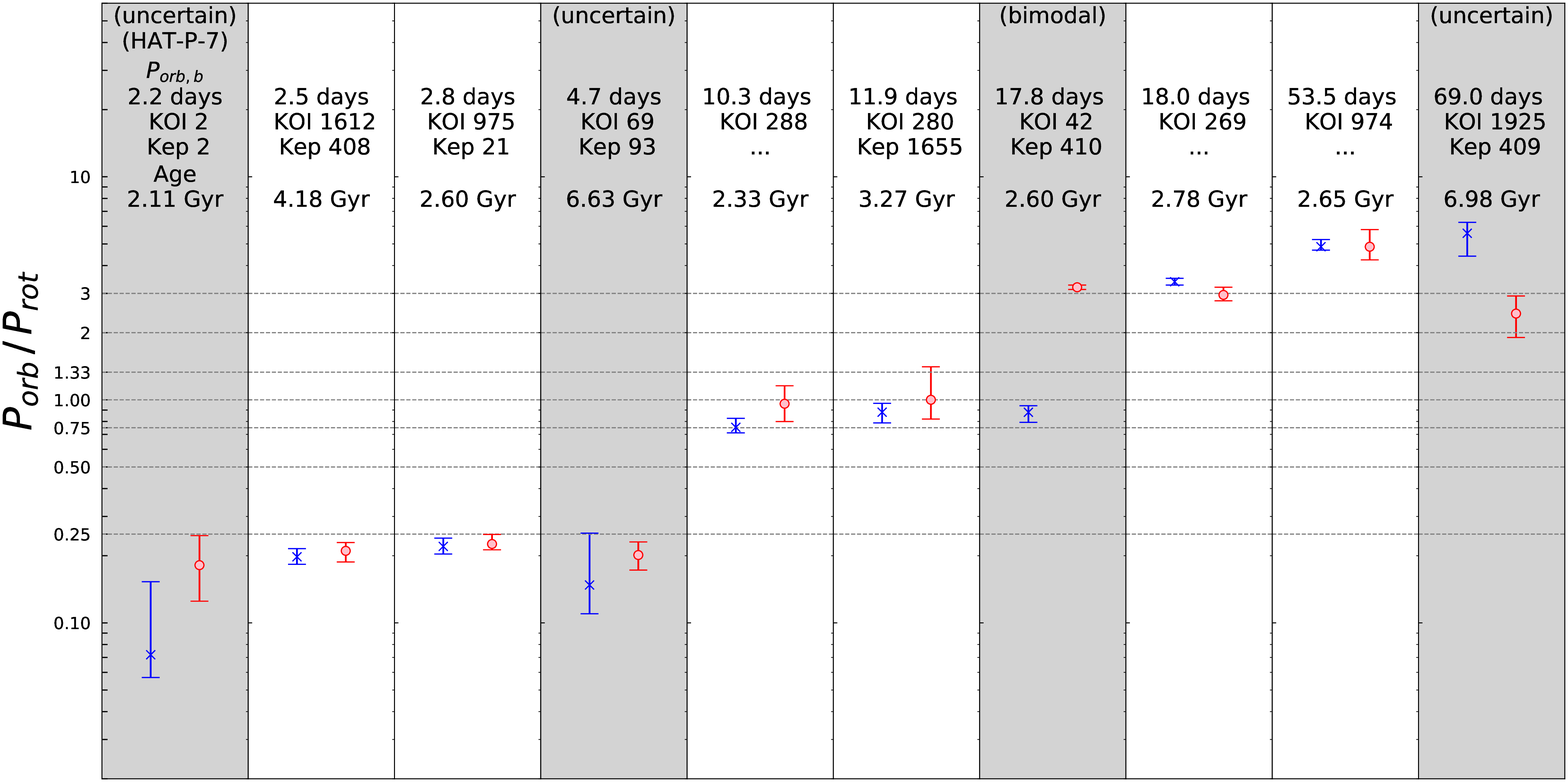}
\caption{Same as Figure \ref{fig:Pratio-multi}, but for
  for single-planet systems.}
\label{fig:Pratio-single}
\end{figure}

Figure \ref{fig:Pratio-multi} shows the trend more specifically and
clearly perhaps, in which the overall spin-orbit architecture for
multi-planetary systems is plotted separately.  Interestingly and
intriguingly, $P_{\rm orb}/P_{\rm rot}$ for seven multi-planetary
systems (except bimodal or uncertain systems shaded as gray) does not
seem to distribute in a homogeneous fashion, but rather preferentially
takes discrete values approximated by simple integer ratios, including
$P_{\rm orb}/P_{\rm rot} =1$.  The most straightforward and bold
interpretation is that those systems are in quasi-spin-orbit resonant
states that have $P_{\rm orb}/P_{\rm rot} \approx n/m$ with $n$ and
$m$ being simple integers.

Figure \ref{fig:Pratio-single} is the same as Figure
\ref{fig:Pratio-multi} but for single-planetary systems (six reliable
systems together with three uncertain and one bimodal systems shaded
as gray).  Apart from the four stars classified as uncertain or
bimodal, a possible (quasi-)spin-orbit resonance is still visible, even though
to a lesser extent than exhibited in Figure \ref{fig:Pratio-multi}.
This may be simply a statistical fluctuation, but may also suggest
that the apparent spin-orbit resonance is somehow related to, or even
enhanced by the orbital resonance in the overall architecture of the
multi-planetary systems.

A strong argument against the interpretation would come from the fact
that the time-scale $\tau_{\rm sync}$ of the spin-orbit
synchronization ($P_{\rm orb, b}/P_{\rm rot} \approx 1$) is
unrealistically long at least in a conventional equilibrium tide model
for a near-circular planetary orbit. Nevertheless we may speculate that
there are a few dynamically stable local minima corresponding to
$P_{\rm orb, b}/P_{\rm rot} \approx n/m$. In the course of the
slow-down of the stellar rotation and/or the planetary migration that
are not directly triggered by the tidal interaction, the star-planet
system may be trapped in one of such quasi-resonant states
temporarily. If this really happens, the corresponding time-scale
could be significantly smaller than $\tau_{\rm sync}$ based on the
mere tidal interaction.  Otherwise the current result would challenge
the existing tidal theories if it is not just a statistical fluke.

\begin{figure}[ht]
  \centering
\includegraphics[width=12cm]{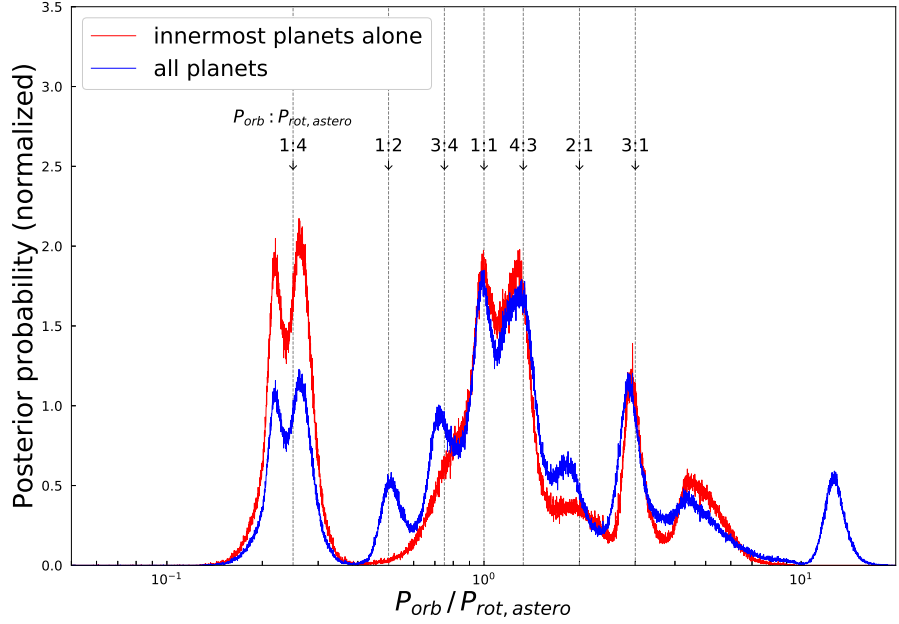}
\caption{Posterior probability density (PPD) for $P_{\rm
      orb}/P_{\rm rot,astero}$ in logarithmic scale.  Red and blue
    lines represent marginalized PPDs based on the
    inner-most planets alone ($N_{\rm planet}=13$) and
    on all the planets  ($N_{\rm planet}=23$). }
\label{fig:PPD}
\end{figure}

Because of the limited number of the planetary systems that allow a
reliable asteroseismic estimate of the stellar rotation period, it is
not easy to provide the statistical significance of the presence of
the possible spin-orbit resonance. Nevertheless we attempt to evaluate
it using the posterior probability density (PPD) for $P_{\rm rot,
  astero}$ obtained by \citet{Kamiaka2018a}. The result is plotted in
Figure \ref{fig:PPD}, which corresponds to the PPD of $x\equiv P_{\rm
  orb}/P_{\rm rot, astero}$ in the logarithmic scale normalized as
\begin{equation}
\int_0^\infty {\rm Prob}(x) {\rm d}\log_{10} x = 1 .
\end{equation}
The red and blue curves indicate the PPD for the inner-most and all
planets, $N_{\rm planet}=13$ and 23, respectively, for the reliable
systems alone. The vertical dotted lines indicate ratios of simple
integers, which may be rather subjective but useful for reference.

We note here that even if the spin-orbit resonance interpretation is
correct, $P_{\rm orb}/P_{\rm rot,astero}$ does not have to coincide
with a ratio of simple integers {\it exactly} as already noted in the
above.  A possible radial differential rotation would make $P_{\rm
  rot,astero}$ slightly different from $P_{\rm rot,true}$.
Furthermore, if the planetary orbit is eccentric, the stellar rotation
velocity would be more likely synchronized towards the planetary
orbital velocity at the pericenter.  Thus $P_{\rm rot,true}$ may well
differ from $P_{\rm orb}$ to some extent.

\subsection{Discussion \label{sec:discussion}}

\begin{figure}[ht]
  \centering
  \includegraphics[width=12cm]{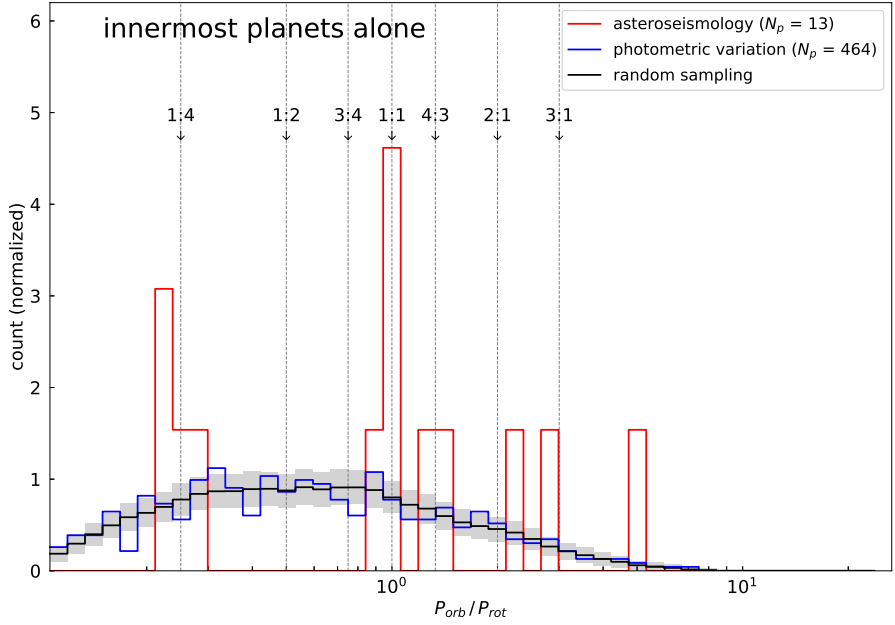}
\caption{Histograms of $P_{\rm orb, b}/P_{\rm rot,astero}$ for our 13
  reliable samples with $P_{\rm rot,astero} \approx P_{\rm
    rot,photo}$.  The ratio is computed from the mean value of $P_{\rm
    rot,astero}$ with the  logarithmically equal bin of 0.05.}
\label{fig:hist}
\end{figure}

We also attempt to compute the chance probability that 3 out of
  13 stars have $0.96<P_{\rm orb, b}=P_{\rm rot}<1.02$; see KOI-262
  (Kepler-50), KOI-280 (Kepler-1655) and KOI-288 in Tables
  \ref{tab:planet_single} and \ref{tab:planet_multi}.  For that
  purpose, we adopt the data-set of the photometric stellar rotation
  period for 464 {\it Kepler} transiting planetary systems compiled by
  \citet{Mazeh2015b}.  Figure \ref{fig:hist} shows the corresponding
  normalized histogram of the ratio of central values of $P_{\rm orb,
    b}$ and $P_{\rm rot, phot}$ in blue.  Then we randomly shuffle
  those values of $P_{\rm orb, b}$ and $P_{\rm rot, phot}$ in the
  systems.  The average and $1\sigma$ region of 1000 sets of random
  sampling are plotted in black line and gray shaded area in Figure
  \ref{fig:hist}. We find that the fraction of a system having
  $0.96<P_{\rm orb, b}/P_{\rm rot, phot}<1.02$ is $p_1=2.12\%$. Thus
  the chance probability that 3 out of 13 stars have $0.96<P_{\rm orb,
    b}/P_{\rm rot, phot}<1.02$ is $p_3={}_3 C_{13}p^3(1-p)^{10}
  \approx 0.22\%$.  If we consider a broader range of $0.9<P_{\rm orb,
    b}/P_{\rm rot, phot}<1.1$ to take account of the associated
  errors, we find that $p_1=6.96\%$ and $p_3=4.68\%$.  While these
  estimates should be still regarded as qualitative, instead of
  quantitative, they are helpful in interpreting the statistical
  significance of the current result.  

\begin{figure}[ht]
  \centering
\includegraphics[width=8cm]{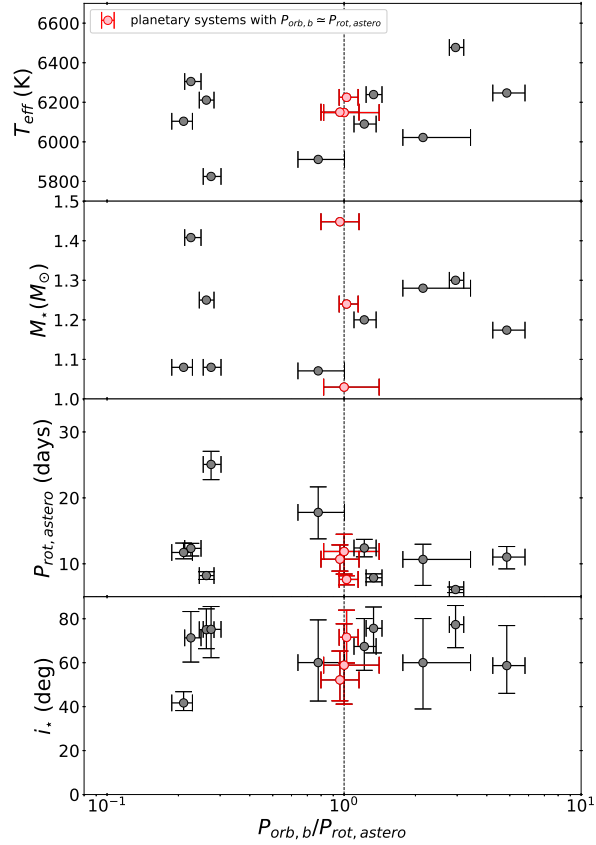}
\caption{$T_{\rm eff}$, $M_\star$, $P_{\rm rot, astero}$ and $\is$ of
    the 13 planetary host stars against $P_{\rm orb,b}/P_{\rm rot,
      astero}$. All the quoted error-bars represent the 68\%
    confidence level.}
\label{fig:star-ratio}
\end{figure}

Figure \ref{fig:star-ratio} plots the stellar parameters for the
  13 reliable systems against $P_{\rm orb, b}/P_{\rm rot,astero}$. Due
  to the limited number of those systems, it is not easy to identify
  any statistical trend, but the three systems with $P_{\rm orb,
    b}/P_{\rm rot,astero} \approx 1$ may have similar effective
  temperature $T_{\rm eff} \approx 6200$K if at all.

\begin{figure}[ht]
  \centering
\includegraphics[width=14cm]{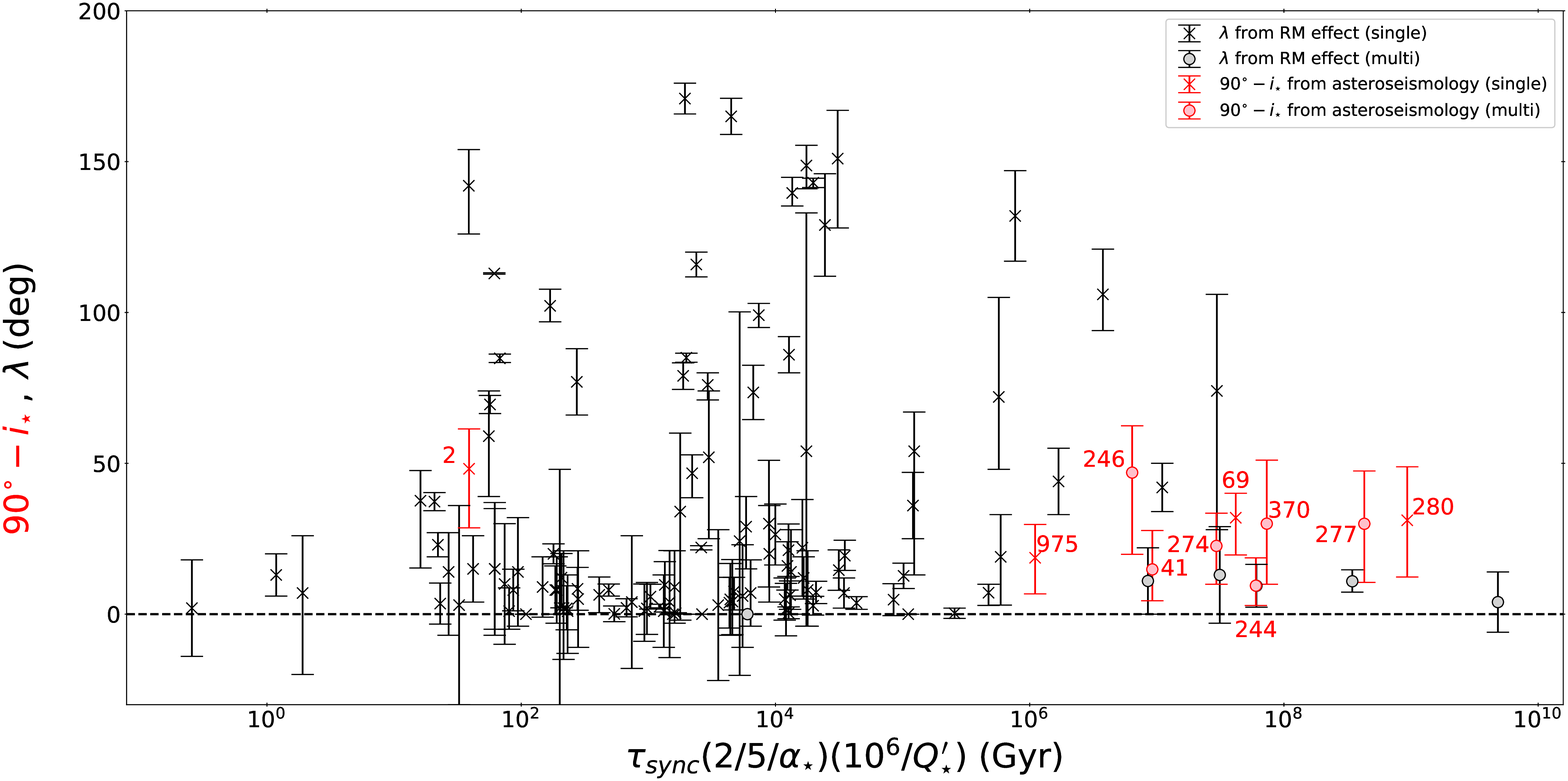}
\caption{Spin-orbit angles $\lambda$ and $90^\circ-\is$ against
  $\tau_{\rm sync}$.  The data for $\lambda$ for 124 transiting
  systems on the basis of the Rossiter-McLaughlin effect in black are
  taken from the compilation by \citet{Southworth2011}. Red symbols
  correspond to 10 systems from our own asteroseismic analysis
  \citep{Kamiaka2018a} with a planetary mass estimate. Crosses and
  circles indicate the single- and multi-planetary systems,
  respectively. The quoted error-bars represent the 68\% credible
  interval.
}
\label{fig:angle-taus}
\end{figure}

Finally we plot the spin-orbit angles $\lambda$ and $90^\circ-\is$
against $\tau_{\rm sync}$ in Figure \ref{fig:angle-taus} \citep[see
  also Figure 12 of][]{Kamiaka2019}.  The black symbols refer to
$\lambda$ from the RM database \citep{Southworth2011}, while the red
symbols are based on our asteroseismic analysis
\citep{Kamiaka2018a}. As discussed in Introduction, the bimodal
distribution of $\lambda$ in Figure \ref{fig:angle-taus} may suggest
the presence of both primordial and realignment channels for the
spin-orbit angle. If those systems around $\lambda \approx 0^\circ$
result from the realignment channel, at least partially, it also points
to stronger tidal interaction because $\tau_{\rm sync}$ of the
conventional equilibrium tide is too long. This may be the same puzzle
that we encounter here in our interpretation of the possible
spin-orbit resonance.

\clearpage

\section{Summary and Conclusion \label{sec:conclusion}}

We have performed asteroseismic analysis of 19 host stars in {\it
  Kepler} transiting planetary systems, and measured their rotation
period $P_{\rm rot, astero}$.  We systematically compared our
measurement against the photometric rotation period $P_{\rm rot, phot}$
estimated in previous literature \citep{Garcia2014,
  Mazeh2015b,Angus2018} and also from our own Lomb-Scargle periodogram
method.

In order to select a robust sample of stars with a reliable rotation
period, we focused on stars that show consistent results between
asteroseismic analysis and the LS periodogram. This turned out to be
particularly important because we found a relatively low-level of
agreement among different published values of $P_{\rm rot, phot}$;
asteroseismology has played a key role in providing an entirely
independent measurement of the stellar rotation.  It is worth noting
that a careful case-by-case examination is necessary if we use
photometric variations (such as the LS periodogram) to derive $P_{\rm
  rot, phot}$.  Indeed, the latitudinal differential rotation, the
size of the star-spots and their typical formation/dissipation
timescales would introduce significant differences between $P_{\rm
  rot, phot}$ and $P_{\rm rot, true}$.

Furthermore, the planet itself induces a photometric modulation that,
if not entirely removed, could be incorrectly identified as the
stellar rotation period. These issues can only be circumvented by
checking results independently with different methods, such as
presented in this study.  Unfortunately, however, measuring the
rotation with seismology requires high quality photometric data, so
that it is difficult to increase the number of reliable stars
significantly.

We found that 13 stars have a strong single peak in the periodogram
and satisfy $0.7<P_{\rm rot,phot}/P_{\rm rot, astero}<1.3$, implying
their rotation period is reliable because $P_{\rm rot,phot}$ and
$P_{\rm rot, astero}$ do not have to be exactly the same due to the
longitudinal and radial differential rotation. The photometric
lightcurves for the remaining six systems exhibit either multiple
peaks (two systems) or no clear peak (four systems) in the
periodogram, and the resulting estimate of $P_{\rm rot,phot}$ is not
reliable. This suggests that the photometrically determined stellar
rotation period needs to be examined carefully on an individual basis.

While the fraction of stars with measured $P_{\rm rot, astero}$ is
relatively small, detailed comparison of their $P_{\rm rot,phot}$ and
$P_{\rm rot,astero}$ is useful in calibrating the reliability of $P_{\rm
  rot,phot}$, and also exploring the spin-orbit architecture of
planetary systems in a robust fashion.

  One straightforward application is the determination of the stellar
inclination $\is$. The asteroseismic estimate of $\is$ for the 13 {\it
  reliable} stars can be more accurate and/or precise with the joint
analysis with $P_{\rm rot, phot}$. A notable example includes
Kepler-408b, the smallest planet known to have a significantly
misaligned orbit with $\is=42^{+5}_{-4}$ degrees \citep{Kamiaka2019}.
The four {\it uncertain} systems, Kepler-2, 68, 93 and 409, may also
be good candidates that host mis-aligned planets.

  Another interesting possibility that we discussed in this paper is the
correlation between the stellar rotation period and the orbital
periods of accompanying planets.  Among the 13 {\it reliable} systems,
we found that three inner-most planets, KOI-262b, 280b, 288b, have
$0.96<P_{\rm orb, b}/P_{\rm rot, astero}<1.02$. On the basis of 464
systems with photometric stellar periods, we estimate that the
corresponding chance probability is ($0.2$ -- $5$) \% depending on the
assumption.

Since the statistical significance for the spin-orbit resonance is
admittedly not so strong, a larger sample of stars would be required
to confirm/refute our current result. Nevertheless, if confirmed, the
(quasi-)spin-orbit resonance points towards a strong tidal interaction
between stars and planets.  This cannot be explained in a conventional
equilibrium tide model.  Due to the limited number of planetary
systems with a reliable stellar rotation period, the interpretation of
the current data is not conclusive yet.  Further investigation on the
basis of the carefully examined photometric variation analysis is of
great importance, which we plan to pursue and report in due course.

\acknowledgments

We thank Adrien Leleu for valuable discussion on the tidal evolution
of the planetary system, and thank the NASA {\it Kepler} team and
KASOC team for making their data publicly available.  We also
acknowledge several critical and constructive comments by Joshua Winn
and Kento Masuda.  The numerical computation was carried out on DALMA
cluster in New York University Abu Dhabi, and PC cluster at Center for
Computational Astrophysics, National Astronomical Observatory of
Japan.  Y.S. gratefully acknowledges the support from Grants-in Aid
for Scientific Research by JSPS No.18H01247.  and by JSPS Core-to-Core
Program ``International Network of Planetary Sciences''.  S.K. is
supported by JSPS (Japan Society for Promotion of Science) Research
Fellowships for Young Scientists (No. 16J03121).  O.B. thanks the
invitation program supported by Research Center of the Early Universe,
the University of Tokyo.


\bibliographystyle{aasjournal}
\bibliography{ref-suto.bib}

\begin{table*}
\renewcommand{\arraystretch}{1.5}
\centering
\caption{Basic stellar properties of 19 planetary systems; $T_{\rm
    eff}$ and $P_{\rm rot, phot}$ denote the effective temperature and
  photometrically-derived rotation period.  The seismically derived
  rotation period, $P_{\rm rot, astero}$, and inclination, $i_{\star,
    \rm astero}$, are estimated using uniform priors, while $i_{\star,
    \rm joint}$ is derived using the photometric rotation period as a
  prior in the seismic analysis. The quoted errors correspond to the
  68\% credible intervals around the median value.  The three systems
  in bold fonts correspond to those with $P_{\rm orb, b}\approx P_{\rm
    rot, astero}$.}
\label{tab:stellar_property}
\begin{tabular}{ccccccc}
\hline
KOI & Kepler ID & $T_{\rm eff}$ & $P_{\rm rot, phot}$ & $P_{\rm rot, astero}$ & $i_{\star, \rm astero}$ & $i_{\star, \rm joint}$ \\
 & & (K) & (days) & (days) & (deg) & (deg)\\
\hline
\multicolumn{7}{l}{{\it Stars with reliable period measurement}} \\
\hline
41 & 100 & 5825 & $27.7_{- 4.2}^{+ 5.0}$ & $25.1_{- 2.3}^{+ 2.0}$ & $75.2_{-12.9}^{+10.4}$ & $77.6_{-11.1}^{+ 8.6}$\\
85 & 65 & 6211 & $ 8.2_{- 0.4}^{+ 0.6}$ & $ 8.2_{- 0.6}^{+ 0.6}$ & $75.0_{- 8.7}^{+ 9.5}$ & $75.4_{- 7.7}^{+ 9.0}$\\
260 & 126 & 6239 & $ 7.2_{- 0.5}^{+ 0.8}$ & $ 7.9_{- 0.6}^{+ 0.6}$ & $75.6_{-11.2}^{+ 9.7}$ & $73.8_{-10.2}^{+10.4}$\\
{\bf 262} & 50 & 6225 & $ 8.1_{- 0.8}^{+ 1.1}$ & $ 7.6_{- 0.8}^{+ 0.6}$ & $71.6_{-11.7}^{+12.3}$ & $75.1_{-10.6}^{+ 9.9}$\\
269 & ... & 6477 & $ 5.3_{- 0.2}^{+ 0.2}$ & $ 6.1_{- 0.5}^{+ 0.4}$ & $77.3_{-10.5}^{+ 8.7}$ & $66.0_{- 5.5}^{+ 7.5}$\\
274 & 128 & 6090 & $13.2_{- 0.9}^{+ 1.1}$ & $12.4_{- 1.3}^{+ 1.3}$ & $67.4_{-10.9}^{+12.7}$ & $71.5_{- 8.4}^{+10.7}$\\
277 & 36 & 5911 & $17.2_{- 1.6}^{+ 1.6}$ & $17.8_{- 4.0}^{+ 3.9}$ & $60.0_{-17.5}^{+19.4}$ & $62.4_{-12.7}^{+16.2}$\\
{\bf 280} & 1655 & 6148 & $13.5_{- 1.2}^{+ 1.6}$ & $11.9_{- 3.4}^{+ 2.6}$ & $58.9_{-17.7}^{+18.8}$ & $68.3_{-11.9}^{+13.3}$\\
{\bf 288} & ... & 6150 & $13.6_{- 1.2}^{+ 0.8}$ & $10.7_{- 1.8}^{+ 2.2}$ & $52.2_{- 9.5}^{+13.1}$ & $67.1_{- 9.6}^{+13.0}$\\
370 & 145 & 6022 & $14.0_{- 1.7}^{+ 1.1}$ & $10.7_{- 3.9}^{+ 2.3}$ & $60.0_{-21.1}^{+20.1}$ & $78.1_{-11.6}^{+ 8.2}$\\
974 & ... & 6247 & $11.0_{- 0.8}^{+ 0.4}$ & $11.0_{- 1.8}^{+ 1.6}$ & $58.7_{-12.6}^{+18.2}$ & $62.1_{- 8.3}^{+12.4}$\\
975 & 21 & 6305 & $12.6_{- 1.0}^{+ 1.0}$ & $12.3_{- 1.2}^{+ 0.8}$ & $71.3_{-11.0}^{+12.0}$ & $75.1_{- 8.8}^{+ 9.8}$\\
1612 & 408 & 6104 & $12.5_{- 1.0}^{+ 1.0}$ & $11.7_{- 1.0}^{+ 1.4}$ & $41.7_{- 3.5}^{+ 5.1}$ & $43.1_{- 2.9}^{+ 3.5}$\\
\hline
\multicolumn{7}{l}{{\it Stars with no clear signal in periodogram}} \\
\hline
2 & 2 & 6389 & $30.6_{-16.2}^{+ 8.1}$ & $12.1_{- 3.2}^{+ 5.5}$ & $41.8_{-13.2}^{+19.6}$ & ...\\
69 & 93 & 5669 & $32.0_{-13.2}^{+11.0}$ & $23.5_{- 3.0}^{+ 3.9}$ & $58.0_{- 8.1}^{+12.3}$ & ...\\
246 & 68 & 5793 & $32.5_{-18.2}^{+ 9.1}$ & $38.0_{-12.8}^{+16.8}$ & $43.1_{-15.5}^{+27.1}$ & ...\\
1925 & 409 & 5460 & $12.4_{- 1.3}^{+ 3.3}$ & $28.3_{- 4.7}^{+ 7.9}$ & $49.8_{- 9.5}^{+16.5}$ & ...\\
\hline
\multicolumn{7}{l}{{\it Stars with bimodal peaks in periodogram}} \\
\hline
42 & 410 & 6273 & $20.3_{- 1.3}^{+ 2.2}$ & $ 5.6_{- 0.1}^{+ 0.1}$ & $83.6_{- 5.2}^{+ 4.4}$ & ...\\
244 & 25 & 6270 & $22.4_{- 1.6}^{+ 3.3}$ & $ 7.8_{- 0.5}^{+ 0.5}$ & $80.6_{- 9.2}^{+ 6.6}$ & ...\\
\hline
\end{tabular}\\
\renewcommand{\arraystretch}{1} {\it References}: $T_{\rm eff}$ is
from NASA Exoplanet Archive
(https://exoplanetarchive.ipac.caltech.edu).
\end{table*}

\begin{table*}
\renewcommand{\arraystretch}{1.5}
\centering
\caption{Properties of single-planetary systems.}
\label{tab:planet_single}
\begin{tabular}{ccccccccc}
\hline
KOI & Kepler ID & $R_{\rm p}$ & $M_{\rm p}$ & $e$ & $a$ & $P_{\rm orb}$ & $P_{\rm orb}/P_{\rm rot, phot}$ & $P_{\rm orb}/P_{\rm rot, astero}$ \\
 &  & ($R_{\oplus}$) & ($M_{\oplus}$) & & (au) & (days) & &\\
\hline
\multicolumn{9}{l}{{\it Stars with reliable period measurement}} \\
\hline
269 & ... & 1.83 & ... & ... & 0.15 & 18.01 & $3.38_{-0.11}^{+0.12}$ & $2.95_{-0.17}^{+0.24}$\\
{\bf 280} & 1655 & 2.21 & 5.0 & ... & 0.10 & 11.87 & $0.88_{-0.09}^{+0.08}$ & $1.00_{-0.18}^{+0.41}$\\
{\bf 288} & ... & 3.04 & ... & ... & 0.10 & 10.28 & $0.75_{-0.04}^{+0.07}$ & $0.96_{-0.16}^{+0.20}$\\
974 & ... & 2.52 & ... & ... & 0.29 & 53.51 & $4.86_{-0.17}^{+0.37}$ & $4.86_{-0.62}^{+0.94}$\\
975 & 21 & 1.64 & 5.08 & 0.02 & 0.04 &  2.79 & $0.22_{-0.02}^{+0.02}$ & $0.23_{-0.01}^{+0.02}$\\
1612 & 408 & 0.82 & ... & ... & ... &  2.47 & $0.20_{-0.01}^{+0.02}$ & $0.21_{-0.02}^{+0.02}$\\
\hline
\multicolumn{9}{l}{{\it Stars with no clear signal in periodogram}} \\
\hline
2 & 2 & 16.9 & 585 & ... & 0.04 &  2.20 & $0.07_{-0.02}^{+0.08}$ & $0.18_{-0.06}^{+0.06}$\\
69 & 93 & 1.6 & 3.2 & ... & 0.05 &  4.73 & $0.15_{-0.04}^{+0.10}$ & $0.20_{-0.03}^{+0.03}$\\
1925 & 409 & 1.19 & ... & ... & ... & 68.96 & $5.58_{-1.18}^{+0.67}$ & $2.44_{-0.53}^{+0.48}$\\
\hline
\multicolumn{9}{l}{{\it Stars with bimodal peaks in periodogram}} \\
\hline
42 & 410 & 2.84 & ... & 0.17 & 0.12 & 17.83 & $0.88_{-0.09}^{+0.06}$ & $3.20_{-0.07}^{+0.07}$\\
\hline
\end{tabular}\\
\renewcommand{\arraystretch}{1} {\it References}: $R_{\rm p}$, $M_{\rm
  p}$, $e$, $a$, and $P_{\rm orb}$ are from NASA Exoplanet
Archive (https://exoplanetarchive.ipac.caltech.edu).
\end{table*}

\begin{table*}
\renewcommand{\arraystretch}{1.5}
\centering
\caption{Properties of multi-planetary systems.}
\label{tab:planet_multi}
\begin{tabular}{ccccccccc}
\hline
KOI & Kepler ID & $R_{\rm p}$ & $M_{\rm p}$ & $e$ & $a$ & $P_{\rm orb}$ & $P_{\rm orb}/P_{\rm rot, phot}$ & $P_{\rm orb}/P_{\rm rot, astero}$ \\
 &  & ($R_{\oplus}$) & ($M_{\oplus}$) & & (au) & (days) & &\\
\hline
\multicolumn{9}{l}{{\it Stars with reliable period measurement}} \\
\hline
41 & 100 & 1.32 & 7.34 & 0.13 & ... &  6.89 & $0.25_{-0.04}^{+0.04}$ & $0.27_{-0.02}^{+0.03}$\\
 &  & 2.20 & ... & 0.02 & ... & 12.82 & $0.46_{-0.07}^{+0.08}$ & $0.51_{-0.04}^{+0.05}$\\
 &  & 1.61 & ... & 0.02 & ... & 35.33 & $1.27_{-0.19}^{+0.22}$ & $1.41_{-0.10}^{+0.14}$\\
85 & 65 & 1.42 & ... & 0.02 & 0.04 &  2.15 & $0.26_{-0.02}^{+0.01}$ & $0.26_{-0.02}^{+0.02}$\\
 &  & 2.58 & 26.6 & 0.08 & 0.07 &  5.86 & $0.72_{-0.05}^{+0.04}$ & $0.71_{-0.05}^{+0.06}$\\
 &  & 1.52 & ... & 0.10 & 0.08 &  8.13 & $1.00_{-0.07}^{+0.06}$ & $0.99_{-0.07}^{+0.08}$\\
260 & 126 & 1.52 & ... & 0.07 & 0.10 & 10.50 & $1.45_{-0.14}^{+0.10}$ & $1.33_{-0.09}^{+0.11}$\\
 &  & 1.58 & ... & 0.19 & 0.16 & 21.87 & $3.02_{-0.30}^{+0.20}$ & $2.77_{-0.19}^{+0.24}$\\
 &  & 2.50 & ... & 0.02 & 0.45 & 100.28 & $13.84_{-1.37}^{+0.92}$ & $12.72_{-0.89}^{+1.09}$\\
{\bf 262} & 50 & 1.71 & ... & ... & 0.08 &  7.81 & $0.97_{-0.12}^{+0.11}$ & $1.02_{-0.07}^{+0.12}$\\
 &  & 2.17 & ... & ... & 0.09 &  9.38 & $1.16_{-0.14}^{+0.14}$ & $1.23_{-0.08}^{+0.15}$\\
274 & 128 & 1.13 & 30.7 & ... & 0.13 & 15.09 & $1.14_{-0.09}^{+0.09}$ & $1.22_{-0.12}^{+0.15}$\\
 &  & 1.13 & 33.3 & ... & 0.17 & 22.80 & $1.73_{-0.14}^{+0.13}$ & $1.84_{-0.18}^{+0.22}$\\
277 & 36 & 1.49 & 4.45 & 0.04 & 0.12 & 13.84 & $0.80_{-0.07}^{+0.08}$ & $0.78_{-0.14}^{+0.23}$\\
 &  & 3.68 & 8.08 & ... & 0.13 & 16.24 & $0.94_{-0.08}^{+0.10}$ & $0.91_{-0.16}^{+0.27}$\\
370 & 145 & 2.65 & 37.1 & 0.43 & ... & 22.95 & $1.63_{-0.12}^{+0.23}$ & $2.15_{-0.38}^{+1.26}$\\
 &  & 4.32 & 79.4 & 0.11 & ... & 42.88 & $3.05_{-0.22}^{+0.42}$ & $4.02_{-0.71}^{+2.36}$\\
\hline
\multicolumn{9}{l}{{\it Stars with no clear signal in periodogram}} \\
\hline
246 & 68 & 2.40 & 6.00 & ... & 0.06 &  5.40 & $0.17_{-0.04}^{+0.21}$ & $0.14_{-0.04}^{+0.07}$\\
 &  & 1.00 & 4.80 & 0.42 & 0.09 &  9.61 & $0.30_{-0.06}^{+0.38}$ & $0.25_{-0.08}^{+0.13}$\\
 &  & ... & 267 & 0.18 & 1.40 & 625 & $19.23_{-4.22}^{+24.63}$ & $16.43_{-5.04}^{+8.28}$\\
\hline
\multicolumn{9}{l}{{\it Stars with bimodal peaks in periodogram}} \\
\hline
244 & 25 & 2.71 & 9.60 & ... & 0.07 &  6.24 & $0.28_{-0.04}^{+0.02}$ & $0.80_{-0.05}^{+0.05}$\\
 &  & 5.20 & 24.60 & 0.01 & 0.11 & 12.72 & $0.57_{-0.07}^{+0.04}$ & $1.64_{-0.09}^{+0.11}$\\
 &  & ... & 89.90 & ... & ... & 123 & $5.48_{-0.70}^{+0.43}$ & $15.83_{-0.90}^{+1.04}$\\
\hline
\end{tabular}\\
\renewcommand{\arraystretch}{1} {\it References}: $R_{\rm p}$, $M_{\rm
  p}$, $e$, $a$, and $P_{\rm orb}$ are from NASA Exoplanet
Archive (https://exoplanetarchive.ipac.caltech.edu).
\end{table*}

\end{document}